\documentclass[aps,jcp,preprint,superscriptaddress]{revtex4-1}
\usepackage{graphicx} % for latex2e
\usepackage{amsmath}
\usepackage{amstext}
\usepackage{amsfonts}
\usepackage{amssymb}
\usepackage{color}

\newcommand{\be}{\begin{equation}}
\newcommand{\ee}{\end{equation}}
\newcommand{\bea}{\begin{eqnarray}}
\newcommand{\eea}{\end{eqnarray}}

\newcommand{\eqa}{\begin{equation}}
\newcommand{\eqz}{\end{equation}}
\newcommand{\eqma}{\begin{eqnarray}}
\newcommand{\eqmz}{\end{eqnarray}}

\usepackage{array}
\newcolumntype{R}[1]{>{\raggedleft  \arraybackslash}p{#1}@{} }
\newcolumntype{C}[1]{>{\centering \arraybackslash}p{#1}@{} }

\newcommand{\cm}{$\text{cm}^{-1}$}
\newcommand{\mx}[1]{\boldsymbol{#1}}

\usepackage{color}

\def\pesname{FullD-2019}

\begin{document}

\title{%
Exact quantum dynamics background of dispersion interactions: 
case study for CH$_4\cdot$Ar in full (12) dimensions \\[0.2cm]
}

\author{Gustavo Avila}
\email{Gustavo$_$Avila@telefonica.net}
\affiliation{Institute of Chemistry, 
ELTE, E\"otv\"os Lor\'and University, 
P\'azm\'any P\'eter s\'et\'any 1/A, 
1117 Budapest, Hungary}

\author{D\'ora Papp}
\email{dorapapp@chem.u-szeged.hu}
\affiliation{
MTA-SZTE Lend\"ulet Computational Reaction Dynamics Research Group, 
Interdisciplinary Excellence Centre and Department of Physical Chemistry and Materials Science, 
Institute of Chemistry, University of Szeged, Rerrich B\'ela t\'er 1, Szeged H-6720, Hungary
}

\author{G\'abor Czak\'o}
\email{gczako@chem.u-szeged.hu}
\affiliation{
MTA-SZTE Lend\"ulet Computational Reaction Dynamics Research Group, 
Interdisciplinary Excellence Centre and Department of Physical Chemistry and Materials Science, 
Institute of Chemistry, University of Szeged, Rerrich B\'ela t\'er 1, Szeged H-6720, Hungary
}

\author{Edit M\'atyus}
\email{matyuse@caesar.elte.hu }
\affiliation{Institute of Chemistry, 
ELTE, E\"otv\"os Lor\'and University, 
P\'azm\'any P\'eter s\'et\'any 1/A, 
1117 Budapest, Hungary}

\date{\today}

\begin{abstract}
\noindent %
A full-dimensional \emph{ab initio} potential energy surface of spectroscopic quality is developed for the 
van-der-Waals complex of a methane molecule and an argon atom.
Variational vibrational states are computed on this surface including all twelve (12)
vibrational degrees of freedom
of the methane-argon complex using the GENIUSH computer program 
and the Smolyak sparse grid method.
The full-dimensional computations make it possible to 
study fine details of the interaction and distortion effects
and to make a direct assessment of the reduced-dimensionality models often used 
in the quantum dynamics study of weakly-bound complexes.
A 12-dimensional (12D) vibrational computation including only 
a single harmonic oscillator basis function (9D) 
to describe the methane fragment 
(for which we use the ground-state effective structure as the reference structure)
has a 0.40~\cm\ root-mean-square error (rms) 
with respect to the converged 12D 
bound-state excitation energies, which is less than half of the rms of the 3D model 
set up with the $\langle r \rangle_0$ methane structure.
Allowing 10 basis functions for the methane fragment in a 12D computation, 
performs much better than the 3D models
by reducing the rms of the bound state vibrational energies to 0.07~\cm.
The full-dimensional potential energy surface correctly describes the dissociation of the system, 
which together with further development of the variational (ro)vibrational methodology 
opens the route to the study of the role of dispersion forces on the excited methane vibrations 
and the energy transfer from the intra- to the intermolecular vibrational modes.
\end{abstract}

\maketitle

\section{Introduction}
\noindent% 
Molecular interactions play an important role in chemistry, biology, and materials science.
Through the many-body construction idea \cite{MuCaFaHuVa84} of the potential energy surface (PES) 
of bulk-phase systems 
the study of molecular interactions is translated to the study of molecular dimers, trimers, and 
perhaps larger (but still small) 
clusters~\cite{QuSu91,QuStSu93,Groenenboom:00,Cencek:08b,WaBo11,MeBaPa13,BaMePa14,Jankowski:15,Gora:14}. 
Small molecular clusters can be studied to a great detail
and precision by high-resolution spectroscopic and quantum chemistry and quantum dynamics
techniques.
A good, `first' description of the quantum dynamical features of molecular complexes 
is provided by the rigid-monomer approximation \cite{HeWoAvMiMo99,SaCsAlWaMa16,SaCsMa17,MeSzSaToCsMa19},
which allows considerable savings both on the PES development and on the quantum dynamics side.
For accounting monomer-flexibility effects through the PES representation,
the application of effective potential energy cuts (for each monomer vibrational state) 
provides an improved representation over the rigid-monomer approach while retaining 
the small number of active vibrational degrees of freedom \cite{JeJaSzJe00}.

At the same time, monomer flexibility `effects' are, of course, 
non-negligible \cite{WaCa17,WaCa18WW}, especially 
for (a) strongly interacting fragments (with strong monomer distortions) 
\cite{AvMa19b,FaCsCz13,WoPaMa12};
(b) higher vibrational excitations; 
(c) monomer vibrational excitations that may correspond to predissociative states
of the complex \cite{ArNOp,FeBa19};
or (d) for symmetry reasons (\emph{i.e.,} degenerate monomer excitations may show a non-trivial coupling
with the intermolecular modes).
A full account of monomer flexibility in complexes of polyatomic molecules
represents a considerable challenge for the current (ro)vibrational methodologies 
due to the large number of vibrational degrees of freedom and 
the typically multi-well character of the potential energy landscape.

A generally applicable, `black-box-type' description of 
molecular systems with multiple-large amplitude motions is truly challenging, 
due to 
    (a) the high-dimensionality of the problem;
    (b) singularities in the kinetic energy operator in the dynamically important region
    of the coordinate space, 
    (c) a common lack of good zeroth-order models, 
    (d) large basis sets and integration grids necessary to converge the results, and thus
    (e) the necessity to attenuate the curse of dimensionality. 
For semi-rigid molecules there have been efficient methods developed in the past \cite{MM2} and 
further major progress has been achieved over the last decade 
\cite{tc-gab1,tc-gab2,AvCa11b,CP1,CP2,betterpr1,betterpr2,MaGoLoCh18,BaRe19}. 
If there is only a single large-amplitude degree of freedom in the system, 
the reaction-path-Hamiltonian \cite{MiHaAd80}
and similar approaches have been successfully used together with 
semi-rigid techniques \cite{BoHuHaCa07,LAUVERGNAT201418}.
There exist efficient, tailor-made approaches developed for particular systems, 
\emph{e.g.,} for molecular complexes \cite{Groenenboom:00,Leforestier:12a}. 
But a general and efficient solution method for systems with
multiple large-amplitude motions remains to be an open problem. 
For this reason molecular systems with multiple-large amplitude motions 
represent a current frontier of research in quantum dynamics.
The present work contributes to this direction.
The family of molecular complexes offer a wide selection of systems with 
a varying number of large- and small amplitude motions, 
varying coupling strengths, singularity patterns, etc., and in this way, 
their study drives methodological developments.

In the present work we focus on the floppy, van-der-Waals complex of
a methane molecule and an argon atom (with twelve vibrational degrees of freedom),
ultimately aiming to reach the predissociative states which belong
to the vibrational excitation of the methane fragment,
within a full-dimensional vibrational treatment.
Due to the weak interactions governing the internal dynamics of the CH$_4\cdot$Ar complex,
powerful approximations could have been introduced in reduced-dimensionality computations, 
including some methane vibrations, to interpret the high-resolution predissociative
spectrum of the complex \cite{HeKoMoWoAv97,HeWoAvMiMo99,MiHeWoAvMo99,WaRoPaWiWoAv03}.
In spite of this earlier experimental and quantum dynamics work (accounting for some
methane flexibility) there is not any full-dimensional (12D) 
potential energy surface available
for this system. Hence, the first part of this article is about the development of an
\emph{ab initio,} near-spectroscopic quality, full-dimensional PES for CH$_4\cdot$Ar. 
The second part reports the first application of 
this PES in vibrational computations including all 12 vibrational degrees of freedom 
using the GENIUSH--Smolyak procedure developed by two of us
in Ref.~\cite{AvMa19a}.
Note that in Ref.~\cite{AvMa19a}, a (3D+9D) PES was used 
(only including kinetic couplings in the Hamiltonian)
in order to be able to test the developed vibrational methodology.
In addition to the development and the first applications of a full-dimensional PES
for CH$_4\cdot$Ar,
we also take the opportunity to test the rigid-monomer (here 3D) approximation(s),
widely used in the study of molecular complexes, 
with respect to the full-dimensional results.

\clearpage
\section{PES development\label{ch:pes}}
\subsection{Computational details}
\subsubsection{Benchmark dissociation energies }
Geometries of the global (GM) and secondary minima (SM) of the CH$_4\cdot$Ar complex are optimized using 
the explicitly-correlated coupled cluster singles, doubles, and perturbative triples electronic structure method, 
CCSD(T)-F12b \cite{AdKnWe07}, with the aug-cc-pVQZ correlation-consistent basis set \cite{Du89}, 
followed by harmonic frequency computations at the same level of theory. The resulting 
equilibrium structures have $C_{3\text{v}}$ point-group symmetry forming three (GM) and one (SM) 
`H-bond(s)'---or, more precisely `H contacts', which modulate the dispersion interaction 
between Ar and CH$_4$. To obtain benchmark dissociation energies ($D_\text{e}$) 
for the GM and SM complexes single-point energy computations are performed at 
the CCSD(T)-F12b/aug-cc-pVQZ geometries: 
CCSD(T)-F12b/aug-cc-pV5Z, CCSD(T) \cite{RaTrPoHG89} and CCSDT(Q) \cite{KaGa05} 
with the aug-cc-pVDZ basis set to obtain post-(T) contributions, and both all-electron (AE) 
and frozen-core (FC) CCSD(T)-F12b/cc-pCVQZ-F12 \cite{WoDu95} to determine core-correlation corrections. 
The FC approach correlates the valence electrons only, whereas in the AE computations the following electrons are 
also correlated: 1s$^2$ for C and 2s$^2$2p$^6$ for Ar. All the \emph{ab initio} computations are carried out 
with the Molpro program package \cite{molpro}, except the CCSD(T) and CCSDT(Q) computations, which 
are performed using the MRCC program \cite{mrcc} interfaced to Molpro. The final benchmark $D_\text{e}$ values are obtained as
\begin{align}
 D_\text{e}(\text{CCSD(T)-F12b/aug-cc-pV5Z}) + \delta[\text{CCSDT(Q)}] + \Delta_\text{core} \; ,  
 \label{eq:eq1}
\end{align}
where
\begin{align}
  \delta[\text{CCSDT(Q)}] = D_\text{e}(\text{CCSDT(Q)/aug-cc-pVDZ}) - D_\text{e}(\text{CCSD(T)/aug-cc-pVDZ})
\end{align}
and
\begin{align}
  \Delta_\text{core} 
    &= D_\text{e}(\text{AE-CCSD(T)-F12b/cc-pCVQZ-F12}) \nonumber \\
    &- D_\text{e}(\text{FC-CCSD(T)-F12b/cc-pCVQZ-F12})\; .
\end{align}

\subsubsection{Full-dimensional PES development}
A full-dimensional analytic \emph{ab initio} PES,
named \pesname\ PES, is developed based on 
15\,995 energy points computed at the CCSD(T)-F12b/aug-cc-pVTZ level of theory at geometries covering 
the configuration space relevant for the interaction between methane and argon. 
Note that previous test computations by one of us showed that the standard augmented and 
F12 correlation-consistent basis sets provide similar accuracy for PES developments \cite{CzSzTe14}.
The geometries used for the 
PES development are generated by isotropically positioning the Ar atom around the methane unit while atoms of 
the equilibrium CCSD(T)-F12b/aug-cc-pVTZ methane structure are also randomly displaced. 
The C--Ar distance is varied between 4 and 20 bohr, and the atoms of methane are displaced in Cartesian coordinates 
within an interval of $[0,0.95]$~bohr. The PES is represented by a polynomial expansion in Morse-like variables of 
the $r_{i,j}$ internuclear distances, $y_{i,j} = \exp(-r_{i,j}/a)$ with $a=2.0$~bohr, and using a compact polynomial basis that is explicitly invariant under permutation of like atoms \cite{BrBo09,BoCzFu11}. 
The highest total polynomial order applied is 7. The total number of the fitting coefficients is 9355. 
A weighted least-squares fit is performed on the energy points, 
where a certain energy $E$ relative to the global minimum has a weight of $(E_0/(E_0 + E)) \times (E_1/(E_1 + E))$  
with $E_0 = 0.05$~hartree and $E_1 = 0.5$~hartree.

\subsection{Results and discussion}
\subsubsection{Benchmark dissociation energies}
{\linespread{1.}
\begin{table}
  \caption{%
  Benchmark dissociation energies ($D_\text{e}$) in \cm\ corresponding to the global (GM) and secondary minimum (SM) 
  structures of the CH$_4\cdot$Ar complex obtained from Eq.~(\ref{eq:eq1}) 
  at CCSD(T)-F12b/aug-cc-pVQZ geometries
  compared to those obtained on the \pesname\ PES developed in this study.
  \label{tab:benchde}
  }
  \scalebox{0.85}{%
  \begin{tabular}{@{}cccc ccc c@{}}
    \cline{1-8}\\[-0.40cm]
    \cline{1-8}\\[-0.40cm]   
       & 
     AVTZ$^\text{a}$ &        
     AVQZ$^\text{b}$ & 
     $\Delta_\text{5Z}$$^\text{c}$  & 
     $\Delta_\text{core}$$^\text{d}$ & 
     $\Delta[\text{CCSDT(Q)}]$$^\text{e}$ & 
     Final$^\text{f}$ & 
     PES$^\text{g}$ \\
    \cline{1-8}\\[-0.40cm]       
    GM & 154.38 & 149.06 & +0.47   &  +1.40 &  +1.90       & 152.83    &  153.13 \\
    SM & 103.46 & 96.79  & $-$0.52 &  +1.22 &  +2.03       & 99.52     &  102.16 \\
    \cline{1-8}\\[-0.40cm]
    \cline{1-8}\\[-0.40cm]
  \end{tabular}
  }
\begin{flushleft}
 $^\text{a}$ $D_\text{e}$(CCSD(T)-F12b/aug-cc-pVTZ)  \\
 $^\text{b}$ $D_\text{e}$(CCSD(T)-F12b/aug-cc-pVQZ)  \\
 $^\text{c}$ $D_\text{e}$(CCSD(T)-F12b/aug-cc-pV5Z)$-D_\text{e}$(CCSD(T)-F12b/aug-cc-pVQZ) \\
 $^\text{d}$ $D_\text{e}$(AE-CCSD(T)-F12b/cc-pCVQZ-F12)$-D_\text{e}$(FC-CCSD(T)-F12b/cc-pCVQZ-F12) \\
 $^\text{e}$ $D_\text{e}$(CCSDT(Q)/aug-cc-pVDZ)$-D_\text{e}$(CCSD(T)/aug-cc-pVDZ) \\
 $^\text{f}$ $D_\text{e}$(CCSD(T)-F12b/aug-cc-pV5Z)$+\Delta_\text{core}+\delta[\text{CCSDT(Q)}]$ \\
 $^\text{g}$ Energy on the PES when the Ar atom was 57~bohr far from the equilibrium structure of methane 
 (the interaction energy is less than 0.001~\cm)
 relative to the corresponding minimum energy of the PES.
\end{flushleft}
\end{table}
}

In Table~\ref{tab:benchde}, we present the benchmark dissociation energies corresponding to the global and 
secondary minimum geometries of the CH$_4\cdot$Ar complex and compare them to the dissociation energies determined 
on the newly developed analytic \pesname\ PES. 
The correction terms listed in Table~\ref{tab:benchde} allow 
for estimating the accuracy of the benchmark dissociation energies. The extremely fast basis set convergence of 
the explicitly-correlated CCSD(T)-F12b method, which is manifested in the $\Delta_\text{5Z}$ corrections of 
only 0.5~\cm, ensures that the CCSD(T)-F12b/aug-cc-pV5Z energy is basis-set-converged within about 0.1--0.2~\cm. 
The correlation of core electrons increases the dissociation energies by around 1.5~\cm, and has an estimated uncertainty 
of 1~\cm. The $\delta[\text{CCSDT(Q)}]$ correlation contributions are also positive values of around 2~\cm\ with a similar estimated uncertainty of 1~\cm. 
(The uncertainty estimates consider basis set effects and post-(Q) contributions.)
Relativistic effects, not taken into account in this work, are supposed to have smaller contribution than core correlation. Taken together, the uncertainty of the final benchmark dissociation energies is estimated to be $\pm 2$~\cm. 
The benchmark $D_\text{e}$ values, as shown in Table~\ref{tab:benchde}, 
are well reproduced on the new PES with 0.3~\cm\ and 2.6~\cm\ differences in the case of the global and 
the secondary minima, respectively.
In the case of the global minimum the above agreement is even better than expected 
due to cancellation of errors as it can be seen from the data of Table I.

\subsubsection{Accuracy of the analytic PES}

{\linespread{1.}
\begin{table}
  \caption{%
    Number of points and root mean square (rms) deviations of the fitting in 
    the chemically interesting energy ranges of the \pesname\ PES relative 
    to its global minimum. \label{tab:pesrms}
  }
\begin{tabular}{@{}c@{\ \ \ \ }c@{\ \ \ \ }c@{}}  
\cline{1-3}\\[-0.4cm]
\cline{1-3}\\[-0.4cm]
  $E_\text{rel}$ range / \cm\  & Number of points &  rms / \cm\ \\
\cline{1-3}\\[-0.4cm]  
  0--11\,000       & 11727 & 0.66 \\
  11\,000--22\,000 &  1073 & 0.90 \\
  22\,000--55\,000 &  1582 & 0.95 \\
\cline{1-3}\\[-0.4cm]
\cline{1-3}\\[-0.4cm]
\end{tabular}
\end{table}
}

\begin{figure}
\includegraphics[scale=0.45]{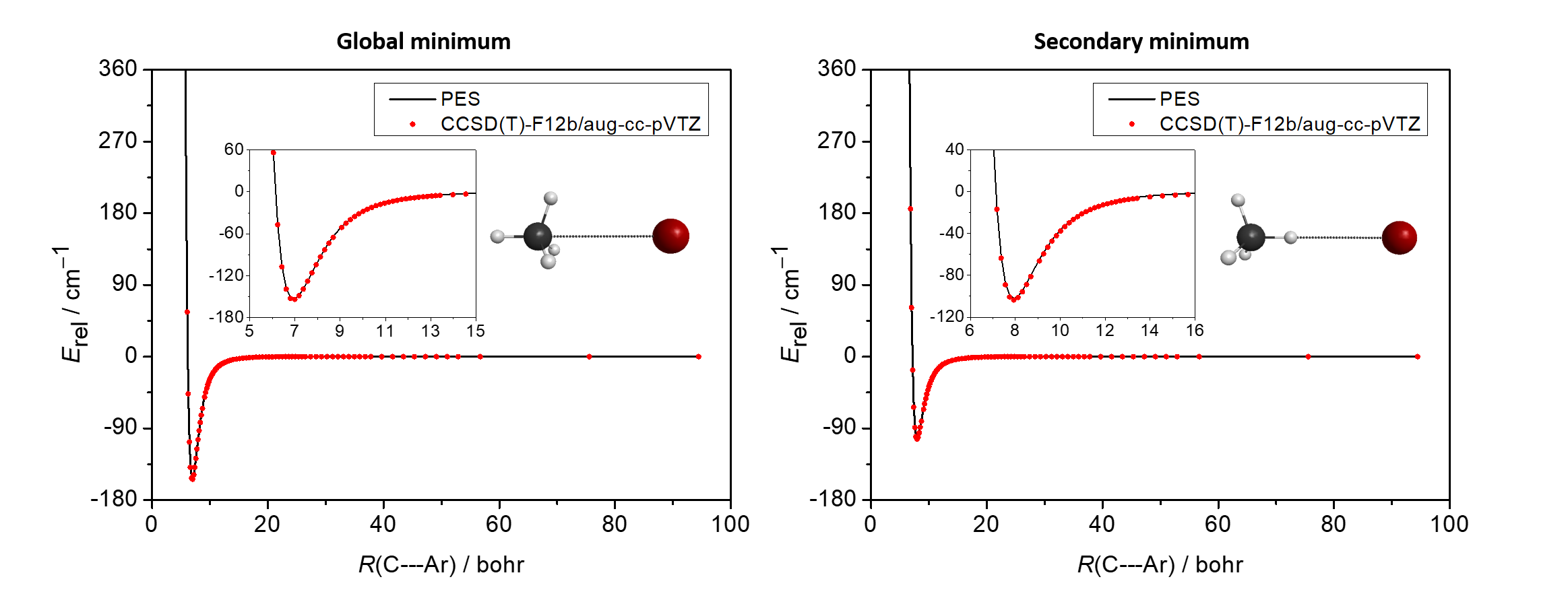}
\caption{%
Potential energy curves along the $C_3$ axes of the global (left panel) and secondary (right panel) minimum structures 
scanning the C--Ar distance of the CH$_4\cdot$Ar complex (the CH$_4$ unit is fixed at its CCSD(T)-F12b/aug-cc-pVTZ 
equilibrium geometry), and showing a comparison between the direct \emph{ab initio} values and 
cuts of the \pesname\ PES.
Insets show the potential well regions and the corresponding equilibrium geometries.
\label{fig:fig1}}
\end{figure}

The newly developed full-dimensional analytic PES of the CH$_4\cdot$Ar complex, \pesname\ PES features extremely 
low root mean square (rms) fitting deviations, listed in Table~\ref{tab:pesrms}, with rms values being lower 
than 1~\cm\ up to 55\,000~\cm\ relative to the global minimum of the PES. 
In accord with these low rms values the one-dimensional energy curves obtained on the PES during
the separation of the Ar atom from methane along the $C_3$ axes of the global and secondary minimum geometries, 
see Figure~\ref{fig:fig1}, show excellent agreement with the \emph{ab initio} energies. As also seen in Figure~\ref{fig:fig1}, 
the asymptotic behavior of the weakly-bound CH$_4\cdot$Ar system is also well described by the PES. 
The asymptotic limits are reached at around 15~bohr from both minima. 
It is worth emphasizing that the fitted PES reproduces the long-range asymptotic behavior of the high-level \emph{ab initio} data without using any 
switching function based on the traditional $1/R^6$ dispersion model.
Figure~2 shows that the structural parameters 
obtained at the minima of the PES agree well with the benchmark CCSD(T)-F12b/aug-cc-pVQZ values. 
The C--Ar distances 
are reproduced on the \pesname\ PES with a difference of 0.003~bohr and 0.029~bohr for the global and the secondary minima, respectively, 
whereas the C--H bond lengths and the H--C--H angles are practically the same as in the benchmark geometry. 
Note that the geometry of CH$_4$ is just slightly perturbed in the minima relative to 
the free CH$_4$ structure; 
the deformation energy at the global minimum is only 0.05~cm$^{-1}$.
The outstanding accuracy of the \pesname\ PES is also strengthened by the dissociation energies corresponding 
to the global and secondary minima reproducing the benchmark values within 3~\cm\ (Table~\ref{tab:benchde}). 
The C--H separation is also scanned on the PES and, as Figure~\ref{fig:fig3} shows, 
\pesname\ PES describes the C--H stretching motion well up to 30\,000~\cm\ relative to the 
global minimum.
Furthermore, the potential scans (Figures 1 and 3) show that the PES function is smooth without any artificial oscillations proving that the large number of fitting parameters does not cause any overfitting problem.

\begin{figure}
\includegraphics[scale=0.5]{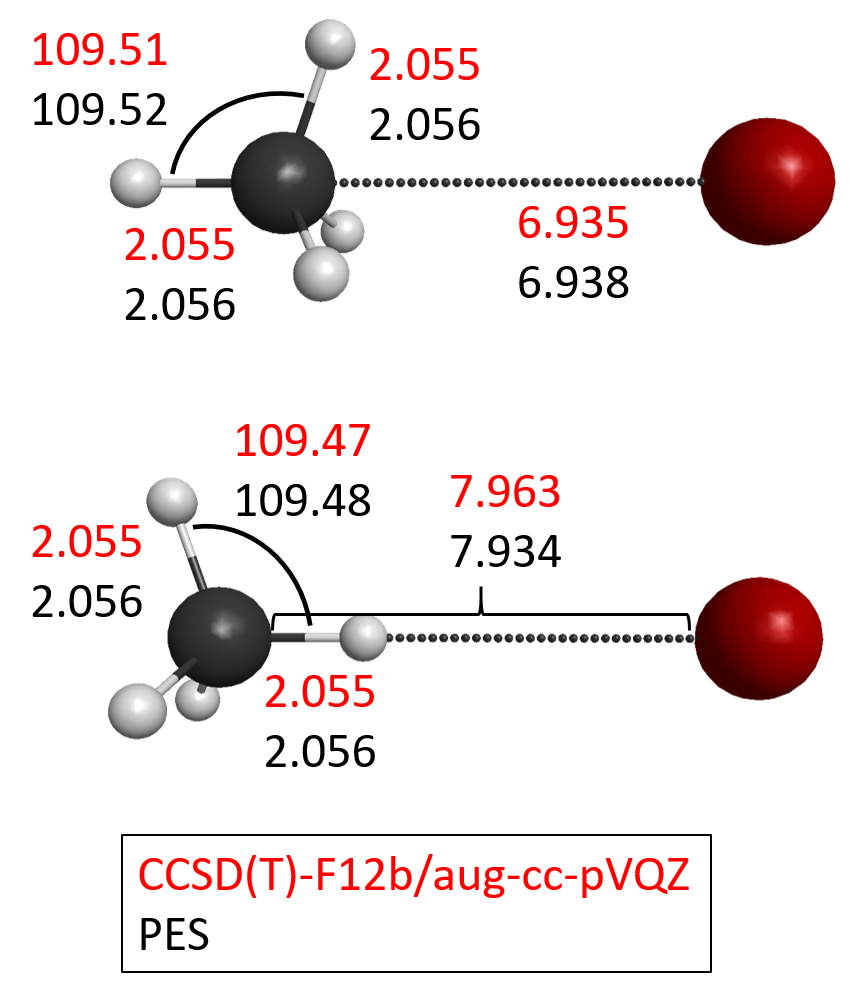}
\caption{%
Geometric parameters of the global (top) and the secondary (bottom) minimum structures of 
the CH$_4\cdot$Ar complex obtained 
at the CCSD(T)-F12b/aug-cc-pVQZ level of theory (red) and on the \pesname\ PES (black). 
Bond lengths are given in bohr and bond angles are given in degree.
\label{fig:fig2}}
\end{figure}

\begin{figure}
\includegraphics[scale=0.6]{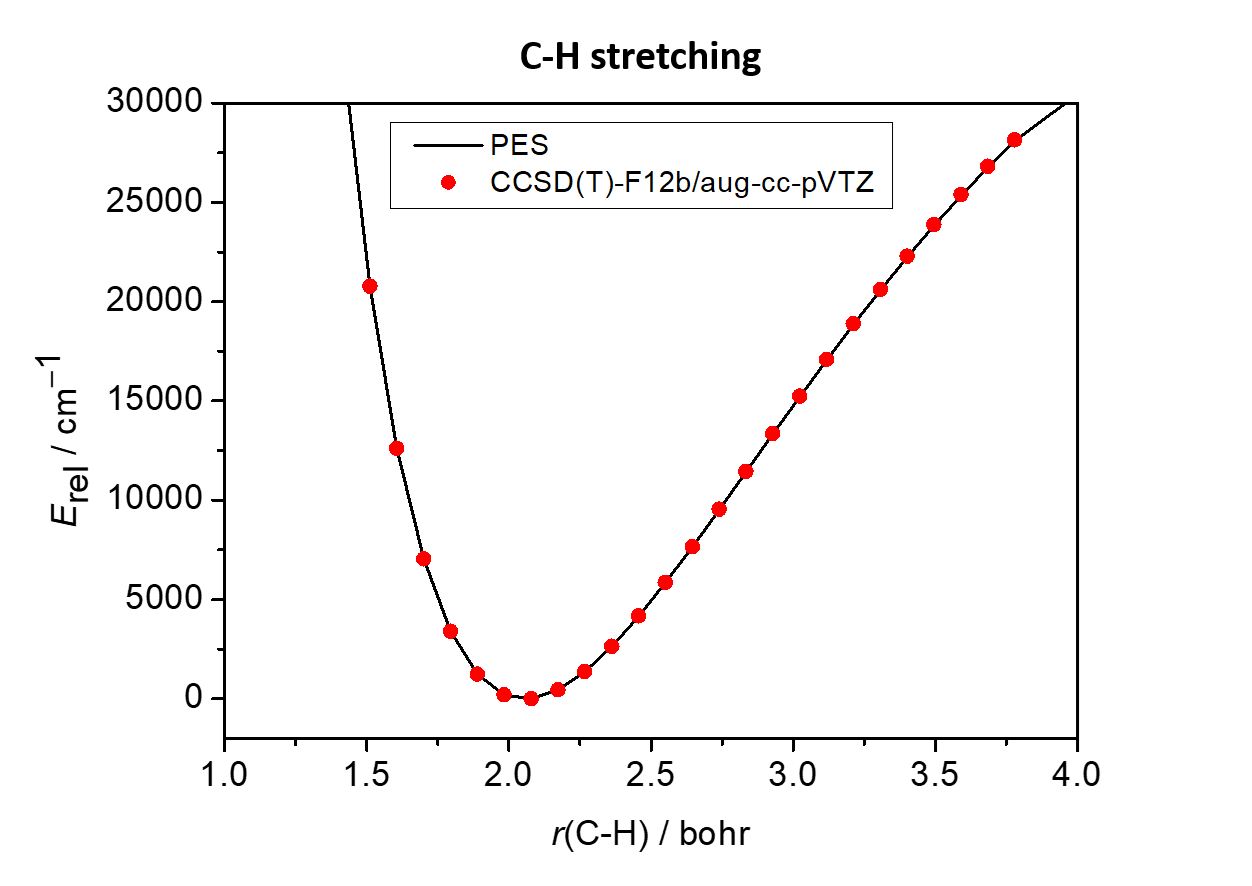}
\caption{%
Potential energy curve along one C--H bond of the CCSD(T)-F12b/aug-cc-pVTZ global minimum geometry 
while the collinear H atom is separated comparing the direct \emph{ab initio} 
values and the \pesname\ PES.
\label{fig:fig3}}
\end{figure}

\clearpage 
\subsection{Asymptotic behavior of the PES and comparison with limiting models}
The \pesname\ PES was fitted to \emph{ab initio} points using a permutationally 
invariant polynomial expansion of Morse variables, $y_{i,j}=\exp(-r_{i,j}/a)$ (with $a=2$~bohr),
which are exponential functions of distances for all atom-atom pairs \cite{BrBo09}.
In the present work, we have included in the fit
polynomials of $y_{i,j}$ up to degree 7.
The advantage of using Morse variables over regular distances 
is that they ensure non-divergent dissociation asymptotes for
large $r_{i,j}$ values. 
At the same time, one might ask whether the exponentially fast decay of the Morse coordinates 
allows us to have a correct description of the intermediate range energetics, 
which, for the present system, is dominated by London dispersion forces, 
commonly described by a potential energy model, which has not an exponential but 
a $1/R^6$-type limiting behavior.

\begin{figure}
\begin{tabular}{@{}c@{\ \ \ }c@{}}
  \includegraphics[scale=0.5]{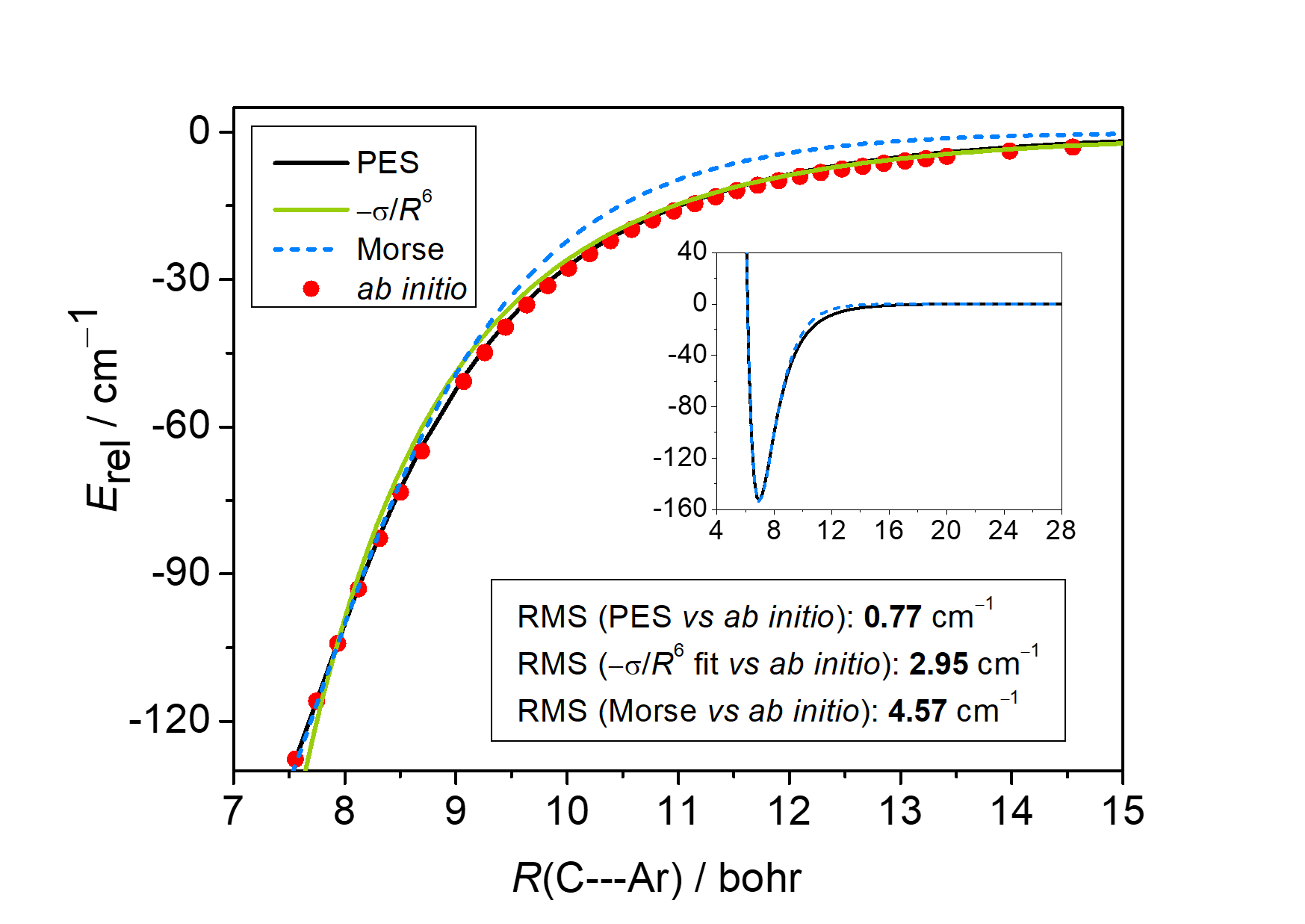} 
\end{tabular}  
\caption{%
Long-range interactions: potential energy representations of CH$_4\cdot$Ar along the dissociation
coordinate showing 
1D cut of the \pesname\ PES, $-\sigma/R^6$ fit 
($\sigma=2.59741\cdot 10^7$~bohr$^6$\ cm$^{-1}$) to
the \emph{ab initio} points in $[7.5, 15.5]$ bohr; 
Morse fit ($153.928\ \text{cm}^{-1}\lbrace 1 - \exp[-0.85 (R / \text{bohr} - 6.95)]\rbrace^2 - 153.928$~cm$^{-1}$); 
and the CCSD(T)-F12b/aug-cc-pVTZ \emph{ab initio} data. 
The RMS values correspond to the data in the [7.5, 15.5]~bohr interval.
The RMS values of the FullD-2019 PES are 0.53 and 0.07~cm$^{-1}$ in [7.5, 95] and [20, 95]~bohr, 
respectively, whereas the corresponding RMS deviations for the $-\sigma/R^6$ fits 
are 1.96 and 0.004~cm$^{-1}$.
\label{fig:fig4}}
\end{figure}

The low rms values (Table~\ref{tab:pesrms}) indicate that the fitting function 
used for the \pesname\ PES had sufficient flexibility to reproduce excellently the \emph{ab initio} energies,
which, of course, automatically capture all `interaction effects'.
To gain more insight in the short, intermediate, and long-range behavior of the system along 
the dissociation coordinate, we compare 
1D cuts in Figure~\ref{fig:fig4}:  the \pesname\ PES,
the Morse potential energy curve,
and the best fit of the $-\sigma/R^6$ model (with $\sigma$ as a constant, fitted parameter)
over the $R\in[7.5,15.5]$~bohr intermediate range.

The Morse potential, $c_0 y^0_R + c_1 y^1_R + c_2 y^2_R$ 
is a second-order polynomial of the $y_R$ Morse variable (defined between
the carbon and the argon atoms), 
and it reproduces excellently the PES valley but it decays too fast to the asymptotic limit 
(Fig.~\ref{fig:fig4}). 
To reproduce well the asymptotic fall, it is necessary to use a higher than second-order 
polynomial in the Morse variable, and we found that a polynomial including monomials
up to the 7th-order, \emph{i.e.,} up to $y^7_R$, in the fitting function of
the \pesname~PES provides an appropriate and automated way to have an excellent overall 
(short-, intermediate-, and long-range) description of the system. 

\begin{figure}
\includegraphics[scale=1.]{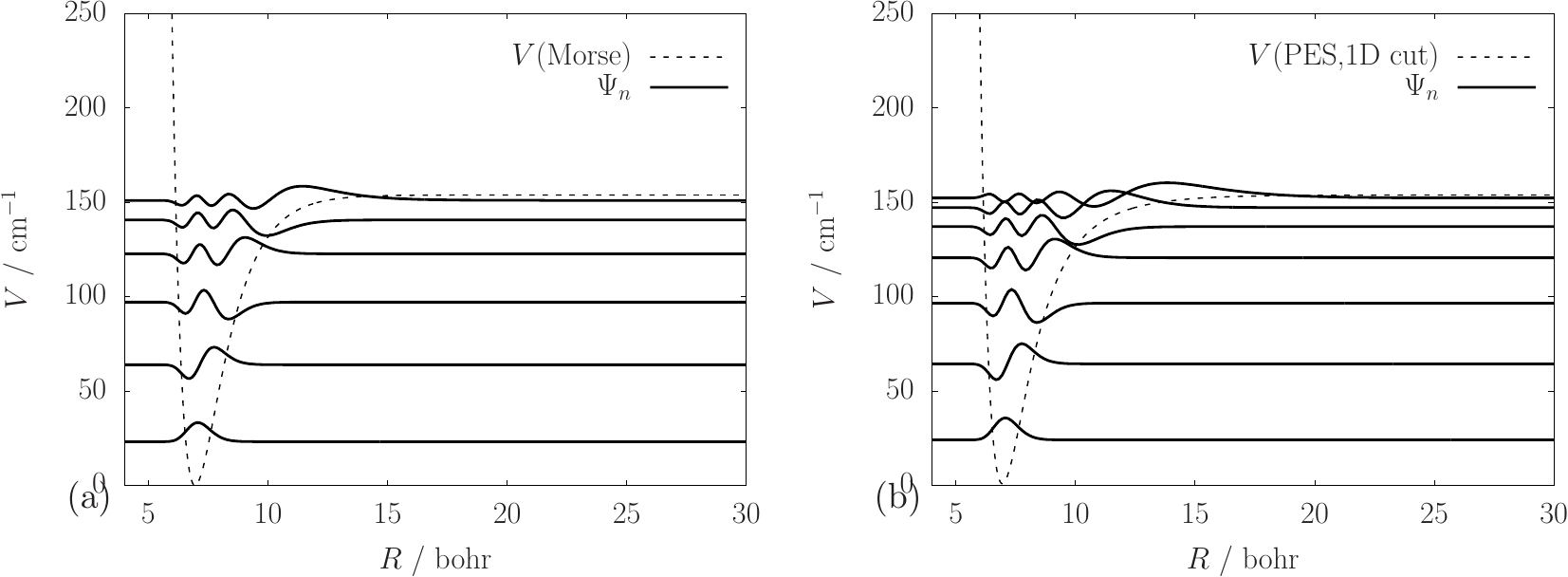}
\caption{%
Vibrational states along the dissociation coordinate (1D model)
using (a) the Morse fit, and (b) the 1D cut of the \pesname\ PES of Figure~\ref{fig:fig4}.
\label{fig:fig5}}
\end{figure}

It is worth pointing out that the difference in the intermediate-range behavior
of the Morse (too fast fall)
and the PES fit (fall similar to the $1/R^6$ dispersion model) 
is manifested also in the vibrational structure (Figure~\ref{fig:fig5}).
The full PES supports an additional bound vibrational state and the highest energy wave function
has a significant amplitude over a much broader range than the highest energy wave function 
corresponding to the Morse fit. 
The significant contribution of higher-order polynomials to the PES representation 
in the long-range asymptotics is observed also in relation with 
using Morse tridiagonal basis functions (Sec.~\ref{ch:interr})
to solve the vibrational problem:
while the low-energy vibrational states can be converged with a small basis set,
the highest-energy bound state requires an excessive number of such functions
indicating that there is an important deviation from the too rapidly decaying Morse character.

%%%%%%%%%%%%%%%%%%%%%%%%%%%%%%%%%%%%%%%%%%%%%%%%%%%%%%%%%%%%%%%%%%%%%%%%%%%%%%%%%%%%%%%%%%%%
%%%%%%%%%%%%%%%%%%%%%%%%%%%%%%%%%%%%%%%%%%%%%%%%%%%%%%%%%%%%%%%%%%%%%%%%%%%%%%%%%%%%%%%%%%%%
%%%%%%%%%%%%%%%%%%%%%%%%%%%%%%%%%%%%%%%%%%%%%%%%%%%%%%%%%%%%%%%%%%%%%%%%%%%%%%%%%%%%%%%%%%%%
\clearpage
\section{Variational vibrational states \label{ch:varvib}}
\noindent %
Using the newly developed \pesname\ PES, 
the bound vibrational states of the methane-argon complex have been computed using 
the GENIUSH--Smolyak approach \cite{AvMa19a}.  
This extension of the GENIUSH program \cite{MaCzCs09,FaMaCs11} 
makes it possible to discard basis functions 
as well as points from the direct product basis and grid, 
using the Smolyak method \cite{tc-gab1,tc-gab2}, 
thereby attenuating the exponential growth 
of the computational cost with the vibrational dimensionality. 
This development makes it possible to solve high-dimensional vibrational problems, 
for which or for, at least, parts of which a good zeroth-order representation 
can be constructed. 

In the case of the CH$_4\cdot$Ar complex, 
a good zeroth-order approximation is obtained for the methane fragment
by using normal coordinates, $(q_1,q_2,\ldots,q_9)$ 
and harmonic oscillator basis functions. 
The relative motion of the fragments is described by spherical polar coordinates, 
$(R,\cos\theta,\phi)$ similarly to Ref.~\cite{AvMa19a}.

The GENIUSH program requires the definition of the internal coordinates (and the body-fixed, BF frame)
by specifying the Cartesian coordinates in the BF frame with respect to the internal coordinates.
The program uses this information to construct the kinetic energy operator (KEO) terms in 
an automated fashion \cite{MaCzCs09}. The usual Cartesian coordinates expression 
of the (generalized) normal coordinates, $q_j\in(-\infty,\infty)$, is
\begin{align}
  r_{i\alpha}
  =
  c_{i\alpha}^{\rm ref} + \sum_{j=1}^{9} l_{i\alpha,j} q_{j} \; ,   
  \label{eq:coormet}
\end{align}
with $i=1(\text{H}),2(\text{H}),3(\text{H}),4(\text{H}),5(\text{C})$ and $\alpha=1(x),2(y),3(z)$.
The $l_{i\alpha,j}$ linear combination coefficients and the 
$c_{i\alpha}^{\rm ref}$ reference structure can be chosen by convenience 
(as a special case, they can be obtained from the harmonic analysis of the PES 
at the equilibrium structure).
In the present work, we chose $c_{i\alpha}^{\rm ref}$ to reproduce
not the equilibrium structure (which could be one of the minima of CH$_4\cdot$Ar 
or the isolated CH$_4$ minimum), but to reproduce the tetrahedral methane
structure for which the C--H distance corresponds to the effective structure of 
the methane zero-point vibration with $\rho:=r_\text{eff}=2.067\,337\,961$~bohr 
\cite{AvMa19a},
\begin{align}
  (\mx{c}^{\text{ref}}_{1})^{\text{T}} &= \frac{\rho}{\sqrt{3}} ( 1, 1, 1)  \nonumber \\
  (\mx{c}^{\text{ref}}_{2})^{\text{T}} &= \frac{\rho}{\sqrt{3}} ( 1,-1,-1) \nonumber \\
  (\mx{c}^{\text{ref}}_{3})^{\text{T}} &= \frac{\rho}{\sqrt{3}} (-1,-1, 1) \nonumber \\
  (\mx{c}^{\text{ref}}_{4})^{\text{T}} &= \frac{\rho}{\sqrt{3}} (-1, 1,-1) \nonumber \\
  (\mx{c}^{\text{ref}}_{5})^{\text{T}} &= (0,0,0) .
\end{align}
This choice accounts for anharmonicity effects on the structure
of the methane fragment already in the coordinate definition.
(Note that here we used an effective structure determined 
in previous work \cite{AvMa19a}
which reproduces the $B_0$ value 
corresponding to the methane PES of Ref.~\cite{WaCa14}.)
Using an effective methane structure corresponding
to the ground-state vibration instead of the equilibrium structure 
for the reference structure of the generalized normal coordinates,
slightly speeds up the convergence of the vibrational energies
with respect to the methane basis. In the `complete basis' limit,
the precise reference structure becomes irrelevant, of course.

The Cartesian coordinates of the argon atom are defined with respect to the 
carbon atom  placed at the origin, using the spherical polar coordinates,
$R\in [0,\infty)$ bohr, $\cos\theta\in [-1,1]$, and $\phi\in[0,2\pi)$,
\begin{align}
  r_{6x} &= R \sin\theta \cos\phi \nonumber \\
  r_{6y} &= R \sin\theta \sin\phi \nonumber \\
  r_{6z} &= R \cos\theta \; . 
  \label{eq:coorar}
\end{align}

In the last step of the coordinate definition, the $r_{i\alpha}$ Cartesian structure, 
Eqs.~(\ref{eq:coormet})--(\ref{eq:coorar}), 
is shifted to the center of mass of the methane-argon complex.
Throughout this work, atomic masses are used, 
$m(\text{H})=1.007\,825\,032\,23$~u, $m(\text{C})=12$~u,
and $m(\text{Ar})=39.962\,383\,123\,7$~u  \cite{NIST}.

In the forthcoming subsections, we first test cuts of the \pesname\ PES
in lower-dimensional vibrational computations.
We report the results of 9D computations carried out for the methane fragment 
(with the argon atom fixed at a large distance),
as well as, observations from 1D ($R$) and 2D ($\cos\theta,\phi$) 
radial and angular model computations are summarized.
The experience gathered from these tests is combined 
to determine the optimal parameters for
the 12D computation, which is presented in the last subsection.
It is important to emphasize that in the final computations 
we include all 12 vibrational degrees of freedom
in the variational vibrational treatment, 
but we chose the coordinates, in particular,
the reference structure of the generalized normal coordinate definition,
so that they provide an excellent description 
for the bound atom-molecule vibrations, 
which are dominated by the methane zero-point state.

\subsection{Isolated methane vibrations \label{ch:metisol}}
The energy levels of the methane molecule were computed on the \pesname\ PES with the argon atom 
fixed at a 30~bohr distance from the center of mass of the CH$_4$ fragment. 
The atom-molecule interaction is (almost) negligible (less than 0.05~\cm) at this separation. 
For the variational computations, we started out from a direct-product basis set of 
harmonic oscillator functions,
$\phi_{n_1}(q_1)\ldots \phi_{n_9}(q_9)$ ($n_i=0,1,\ldots, i=1,\ldots 9$), 
which was pruned according to the simple condition
$n_1+\ldots +n_9\leq b$. 
An integration grid (much) smaller than the na\"ive direct-product grid 
was defined using the Smolyak scheme \cite{tc-gab1,tc-gab2}. 
In short, the grid points were chosen to integrate exactly the matrix of 
the identity and also polynomials of up to a maximum degree of 5 
with the basis functions included in the pruned basis set \cite{AvMa19a}.
The pruning parameter, the size of the basis, and the size of the Smolyak grid are listed 
in Table~\ref{tab:metpar}.
The convergence rate with respect to the basis and grid size
as well as benchmark results for the vibrational energies of CH$_4$
are shown in Table~\ref{tab:isolmet}. 

Using the $b=10$ basis-pruning parameter,
the energies are converged within 0.01~\cm\ up to (and including) the pentad of CH$_4$. 
The 9690.62~\cm\ zero-point vibrational energy (ZPVE) on the \pesname\ PES 
is in good agreement with the 9691.56~\cm\ value corresponding to the T8 
force field of Schwenke and Partridge \cite{ScPa01}.
The root-mean-square (rms) deviation of the converged vibrational excitation energies 
with respect to their
counterparts deduced from experiments \cite{NiReTy11} is 2.88~\cm, which is excellent given that
this is a purely \emph{ab initio} PES, which was developed not specifically for 
an isolated methane molecule but for the methane-argon complex. Note that these `isolated methane'
energies were obtained using the \pesname\ PES with the argon atom fixed at a large
distance from the methane molecule. 

Assessment of smaller basis sets (smaller $b$ values) is important for planning the 12D 
computations. The bound states of the CH$_4\cdot$Ar complex are  dominated
by the zero-point state of methane, hence, $b=3$ should be an excellent compromise
for computing the intermolecular (atom-molecule) states accurately. 
The computation of predissociative states corresponding to 
excited vibrational states of methane will require at least
$b=6$--7, which assumes further development of the vibrational
methodology.

{\linespread{1.}
\begin{table}
\caption{%
  Basis set and integration grid parameters used to describe the methane fragment.
  \label{tab:metpar}
}
\begin{tabular}{@{}c@{\ \ \ }c@{\ \ \ }r@{\ \ \ }r@{}}
\cline{1-4}\\[-0.4cm]
\cline{1-4}\\[-0.4cm]
\multicolumn{1}{c}{$b$$^\text{a}$} & 
\multicolumn{1}{c}{$H$$^\text{b}$} & 
\multicolumn{1}{c}{$N_\text{bas}$$^\text{c}$} & 
\multicolumn{1}{c}{$N_\text{Smol}$$^\text{d}$} \\
\cline{1-4}\\[-0.35cm]
 0 &  11$^\text{e}$  &        1 &     163$^\text{e}$ \\
 1 &  12$^\text{e}$  &       10 &     871$^\text{e}$ \\
 2 &  13  &       55   &      3\,481 \\
 3 &  14  &      220   &     11\,833 \\
 4 &  15  &      715   &     35\,929 \\
 5 &  16  &     2\,002 &     97\,561 \\
 6 &  17  &     5\,005 &    241\,201 \\
 7 &  18  &    11\,440 &    556\,707 \\
 8 &  19  &    24\,310 & 1\,202\,691 \\
 9 &  20  &    48\,620 & 2\,440\,227 \\
10 &  21  &    92\,378 & 4\,718\,595 \\
\cline{1-4}\\[-0.4cm]
\cline{1-4}\\[-0.4cm]
\end{tabular}
\begin{flushleft}
  $^\text{a}$ % 
    Basis pruning condition, $n_{1}+\ldots+n_{9}\leq b$. \\
  $^\text{b}$ % 
    Grid pruning condition, $i_{1}+\ldots+i_{9}\le H$
    (for details, see for example, Ref.~\cite{AvMa19a} and references therein). 
    We chose $H=D-1+b+3$ (here $D=9$), to integrate exactly not only the overlap 
    but also polynomials of a maximum degree of 5     
    with all basis functions included in the pruned basis set. \\
  $^\text{c}$ %  
    The number of basis functions in the pruned basis set is $N_\text{bas}=(b+9)!/(b!9!)$. \\
  $^\text{d}$ %  
    The number of points in the Smolyak grid corresponding to the selected $H$ value. \\
  $^\text{e}$ %
    When using the \pesname\ PES with $b=0$ ($b=1$), 
    we observed that $H=11$ ($H=12$) is not sufficient to recover the correct degeneracy
    of the methane vibrations (especially the E states were affected). So, in the end,
    we used $H=12$ ($H=13$) and $N_\text{Smol}=871$ ($N_\text{Smol}=3481$)
    for $b=0$ ($b=1$).
\end{flushleft}
\end{table}
}

{\linespread{1.}
\begin{table}
  \caption{%
    Convergence of the zero-point and vibrational excitation energies, in \cm, up to and including the pentad of 
    CH$_4$ with respect to the (pruned) basis set size using the GENIUSH--Smolyak approach~\cite{AvMa19a}
    and the \pesname\ PES with an argon-methane distance fixed at $R=30$~bohr.
    The benchmark energies corresponding to this PES are given in the $\tilde\nu(b=10)$ column.
    \label{tab:isolmet}
   }
\scalebox{0.9}{%
\begin{tabular}{@{}cc @{\ \ \ }c@{\ \ \ }c@{\ \ \ }c@{\ \ \ }c@{\ \ \ }c@{\ \ \ }c@{\ \ \ }c@{\ \ \ }c@{\ \ \ }c@{\ \ \ }c@{\ \ \ }c c@{\ \ \ \ \ }c@{}}
\cline{1-15}\\[-0.4cm]
\cline{1-15}\\[-0.4cm]
\multicolumn{1}{c}{$\Gamma$$^\text{a}$} &
\multicolumn{1}{c}{Label$^\text{b}$} &
\multicolumn{1}{c}{$\Delta_0$$^\text{c}$} &
\multicolumn{1}{c}{$\Delta_1$$^\text{c}$} &
\multicolumn{1}{c}{$\Delta_2$$^\text{c}$} &
\multicolumn{1}{c}{$\Delta_3$$^\text{c}$} &
\multicolumn{1}{c}{$\Delta_4$$^\text{c}$} &
\multicolumn{1}{c}{$\Delta_5$$^\text{c}$} &
\multicolumn{1}{c}{$\Delta_6$$^\text{c}$} &
\multicolumn{1}{c}{$\Delta_7$$^\text{c}$} &
\multicolumn{1}{c}{$\Delta_8$$^\text{c}$} &
\multicolumn{1}{c}{$\Delta_9$$^\text{c}$} &
\multicolumn{1}{c}{${\tilde\nu}(b=10)$$^\text{c}$} & 
\multicolumn{1}{c}{$\delta$$^\text{d}$} &
\multicolumn{1}{c}{${\tilde\nu}_\text{exp}$$^\text{e}$} \\
\cline{1-15}\\[-0.4cm]
 $A_{1}$ & 0000   & 43.79 &  42.42 &        40.06 &       1.79 &       0.60 &       0.52 &       0.04 &       0.01 &       0.01 &       0.00 &    9690.62  &  --  & -- \\
 $F_{2}$ & 0001   &   --  &  9.61  &         7.68 &      41.19 &       2.14 &       0.27 &       0.54 &       0.06 &       0.01 &       0.01 &    1310.60  & 0.16 & 1310.76\\
 $E$     & 0100   &   --  &  5.76  &         5.55 &      39.20 &       1.76 &       0.18 &       0.50 &       0.04 &       0.01 &       0.01 &    1531.47  & 1.86 & 1533.33\\
 $A_{1}$ & 0002   &   --  &  --    &        37.74 &      56.88 &      43.69 &       4.60 &       1.24 &       0.64 &       0.11 &       0.02 &    2586.02  & 1.02 & 2587.04 \\
 $F_{2}$ & 0002   &   --  &  --    &        29.71 &      54.22 &      45.63 &       3.86 &       1.12 &       0.66 &       0.10 &       0.02 &    2613.61  & 0.65 & 2614.26\\
 $E$     & 0002   &   --  &  --    &        19.70 &      51.49 &      45.00 &       3.13 &       0.94 &       0.62 &       0.08 &       0.01 &    2623.93  & 0.69 & 2624.62\\
 $F_{2}$ & 0101   &   --  &  --    &        23.55 &      51.84 &      40.55 &       3.20 &       0.97 &       0.56 &       0.07 &       0.01 &    2828.08  & 2.24 & 2830.32 \\
 $F_{1}$ & 0101   &   --  &  --    &        17.14 &      49.64 &      43.22 &       2.87 &       0.87 &       0.59 &       0.07 &       0.01 &    2844.38  & 1.70 & 2846.08 \\
 $A_{1}$ & 1000   &   --  & 89.70  &        34.44 &      66.84 &      10.68 &       1.63 &       1.38 &       0.27 &       0.05 &       0.02 &    2912.36  & 4.12 & 2916.48\\
 $F_{2}$ & 0010   &   --  & 107.24 &        38.27 &      69.89 &      11.81 &       1.76 &       1.50 &       0.29 &       0.05 &       0.03 &    3014.47  & 5.02 & 3019.49 \\
 $A_{1}$ & 0200   &   --  &  --    &        15.09 &      49.17 &      41.73 &       2.73 &       0.82 &       0.56 &       0.06 &       0.01 &    3059.25  & 4.40 & 3063.65\\
 $E$     & 0200   &   --  &  --    &        13.75 &      47.58 &      41.39 &       2.48 &       0.78 &       0.55 &       0.06 &       0.01 &    3061.06  & 4.08 & 3065.14\\
\cline{1-15}\\[-0.4cm]
 rms     &        &       &        &              &            &            &            &            &            &            &            &             & 2.88 & \\
\cline{1-15}\\[-0.4cm]
\cline{1-15}\\[-0.4cm]
\end{tabular}
}
\begin{flushleft}
$^\text{a}$
  Label of the irreducible representation of the $T_\text{d}$ point group of methane. \\
$^\text{b}$
  `$n_1n_2n_3n_4$' normal mode label. \\
$^\text{c}$
    Deviation from the $\tilde\nu(b=10)$ benchmark value, 
    $\Delta_k=\tilde\nu(b=k)-\tilde\nu(b=10)$. \\
$^\text{d}$
  $\delta=\tilde\nu_\text{exp}-\tilde\nu(b=10)$. \\
$^\text{e}$
    Vibrational energies deduced from experiments are taken from Ref.~\cite{NiReTy11}.
\end{flushleft}
\end{table}
}

\clearpage
\subsection{Intermolecular radial representation \label{ch:interr}}     
There are several possibilities to describe the vibrational motion 
along the methane-argon distance. One can use
$\mathcal{L}_{n}^{(\alpha)}$ generalized Laguerre basis functions 
(with $\alpha=2$) \cite{FeMa19,ArNOp}
or a Morse tridiagonal basis set. 
The Laguerre basis set may be a better choice for computing predissociative states, 
whereas the Morse tridiagonal basis 
set offers a more compact alternative for bound states. 
In the present work, we used
the Morse tridiagonal basis set parameterized with the $D=150$~cm$^{-1}$, $\alpha=0.65$, 
and $\gamma=0.00033$ values \cite{WeCa92,doi:10.1063/1.451775,doi:10.1063/1.444316,AvMa19a,AvMa19b}, 
which gave a good Morse fit 
to the cut of the \pesname~PES at the equilibrium (global minimum) structure of 
all other coordinates.
Since CH$_{4}\cdot$Ar is an isotropic complex, 
this radial basis 
is expected to perform well over the entire range of the angular coordinates.
The convergence tests suggest that 
13 Morse functions with 15 quadrature points for $R$
allow us to converge the 3D($R,\cos\theta,\phi$)  and 12D bound-state energies 
within 0.01~\cm.

\subsection{Intermolecular angular representation\label{ch:intera}}
For the $\cos\theta$ coordinate, we use sin-cot-DVR 
(DVR, discrete variable representation) 
basis functions and points \cite{SCHIFFEL2010118}, 
while Fourier basis functions are used for the $\phi$ angle.
% {\colred
Test computations suggest that 23 sin-cot-DVR 
 functions for $\cos\theta$ and 
 21 Fourier functions with 24 quadrature points for $\phi$ 
 will be sufficient to converge the 3D($R,\cos\theta,\phi$) and 12D vibrational excitation energies 
 better than 0.01~\cm. 
% }

\subsection{%
  Full-dimensional (12D) vibrational states and comparison with 3D models 
  \label{ch:12d}
}
All vibrational bound states of the CH$_4\cdot$Ar complex on the newly developed \pesname\ PES 
are listed in Table~\ref{tab:vib}.
The intermolecular basis set corresponding to the $(R,\cos\theta,\phi)$ coordinates 
is sufficiently large to converge all vibrational
excitation energies better than 0.01~\cm. 
In order to find the smallest necessary intramolecular methane basis set 
(characterized with the $b$ basis-pruning parameter), 
we have carried out 12D computations 
with an increasing basis set size on the methane fragment
corresponding to the $b=0,1,2,$ and 3 value.
Concerning the ZPVE, we think that the $b=3$ 12D result is ca. $1-2$~\cm\ 
larger (our approach is nearly variational) than the exact result, similarly
to the $b=3$ ZPVE value of isolated methane (Table~\ref{tab:isolmet}).
Concerning the intermolecular vibrational excitation energies, 
we could efficiently rely on the cancellation of error in 
the relative vibrational energies, 
and thus, a rather small methane basis set was sufficient to achieve 
the 0.01~\cm\ convergence goal for the excitation energies. 

It is interesting to consider the 
convergence of the excitation energies with respect to $b$.
The 12D computation with $b=0$, which corresponds to 
a single(!) basis function on the methane fragment, 
has an rms error of 0.40~\cm. This rms value 
is ca.~half of the rms error of a well-converged 3D computation 
imposing rigorous geometrical constraints with 
an $\langle r\rangle_0$ methane structure \emph{(vide infra)}. 
A 12D computation with $b=1$, which includes 10 basis functions 
for the methane fragment,
has an rms of 0.07~\cm. Finally, our 0.01~\cm\ convergence goal 
is achieved for the $b=2$ and $b=3$ pruning parameter values.

Concerning the computational cost,
the $b=0$ and $b=1$ computations took 10 and 20~hours (using 20 processor cores)
and required 6 and 8 GB of memory, respectively. 
The 12D $b=3$ ($b=2$) computations 
took 42 (13) days on 50 (20) cores and required 80 (30)~GB of memory.
As it was indicated already in the footnote to Table~\ref{tab:metpar},
we had to use a larger grid size for $b=0$ and 1, than we had originally anticipated, 
which slightly increased the cost of the computation.
Furthermore, the condition number of the Hamiltonian matrix in the current representation
is very large (due to the application of sin-cot-DVR basis functions, there are grid points
which are very close to the singularities of the KEO), which implies an increased number of 
Lanczos iteration steps. We anticipate reduction of the computational cost
with further developments.

\vspace{0.5cm}
Table~\ref{tab:vib} also shows the result of 3D, 
rigid-monomer computations, in which 
only the $R,\cos\theta,$ and $\phi$ degrees of freedom were treated as active coordinates.
The `3D($\langle r\rangle_0$)' column corresponds to reduced-dimensionality results
in which rigorous geometrical constraints were imposed on the methane's structure
(referred to as `the reduction in the Lagrangian' or 
`reduction in the $\boldsymbol{g}$ matrix' in Ref.~\cite{MaCzCs09}
and constructed automatically in GENIUSH).
The methane was fixed at a regular tetrahedral structure with
$\langle r_{\text{C--H}} \rangle_0$, which we calculated 
as the expectation value of the C--H distance using 
the isolated methane's ground-state wave function on the present PES. 
The vibrational excitation energies of this 3D model 
have a relatively large, 0.93~\cm, rms with respect to the converged 12D result. 
Furthermore, this 3D model (erroneously) predicts an additional,
triply degenerate, bound state below the dissociation asymptote,
which can be explained by the slightly different $B_0$ value corresponding
to this model.

In the `3D($\langle B\rangle_0$)' column, 
we report the bound vibrational energies obtained
with an `adjusted' 3D model.
While using the $\langle r_\text{C--H}\rangle_0$ value for defining 
the 3D cut of the PES,
we adjusted the C--H distance in the KEO to reproduce the 
$\langle B\rangle_0$ effective rotational constant of this PES in
2D coupled-rotor computations \cite{SaCsMa17,FeMa19}. 
This model reproduces the correct
number of bound states and has a smaller, 0.32~\cm\ rms, than the 3D model with 
the rigorous geometrical constraints.

In relation with these 3D models, we conclude
that a 12D computation performed with a single 9D basis function 
for methane ($b=0$) is on par with the 3D($\langle B \rangle_0$) model. 
If we allow only 10 functions for the methane fragment ($b=1$) in the 12D computation,
the 12D result clearly outperforms the 3D excitation energies, 
without increasing the computational cost dramatically.

\vspace{0.5cm}
In order to rationalize these numerical observations, we may distinguish between
`static' and `dynamical' contributions from the methane's vibrations
on the atom-molecule energy levels. 
The static contribution is due to the fact that the isolated molecule's 
effective (average) structure,
due to anharmonicity of the methane's vibrations, 
is different from the equilibrium structure.
In 3D computations, this effect is accounted for by 
fixing the methane's structure at an effective structure instead of the 
equilibrium structure.
In 12D computations, we have `built in' this static effect 
in the coordinate definition (using generalized normal coordinates) 
in order to speed up convergence with respect to the methane's basis size.

The dynamical contribution is due to the coupling of the methane's vibrations 
with the intermolecular dynamics. 
This dynamical coupling, which is often small but non-negligible, 
requires a full-dimensional treatment.
In the case of the methane-argon complex, 
we observe that using only the ground and all singly excited
(9D) harmonic oscillator functions capture almost all dynamical effects,
but well-converged excitation energies assume
at least $220$ (9D) harmonic oscillator functions,
corresponding to the $n_1+n_2+\ldots + n_9 \leq 3$ pruning condition, 
for the methane fragment.

From a numerical point of view, 
it is necessary to mention that the 12D and 3D results are close, within 1-2~cm$^{-1}$, 
to the vibrational states reported in Ref.~\cite{HeWoAvMiMo99} (Table~IV)
on a 3D PES developed using symmetry-adapted perturbation theory and an effective
(ro)vibrational Hamiltonian some 20 years ago. 
The present work reports a fully \emph{ab initio} quantum dynamics study of the system in full dimensionality 
and its further extension to predissociative states may reveal larger deviations from 
the effectively designed approach of Refs.~\cite{HeKoMoWoAv97,HeWoAvMiMo99}. 
There are also experimental results in the predissociative range \cite{MiHeWoAvMo99,WaRoPaWiWoAv03}, 
which will make the comparison
more interesting.
Furthermore, the methodology recently developed and used in the present work is not 
designed specifically for very weakly interacting atom-molecule complexes, but it can be used 
for a greater variety of molecular systems and it fits in a series of recent efforts made for 
a systematic development of general `black-box-type' (ro)vibrational quasi-variational methods,
applicable to molecular complexes or any other molecular systems,
of high vibrational dimensionality and with multiple large-amplitude motions.

{\linespread{1.}
\begin{table}
\caption{%
 All bound-state vibrational energies, $\tilde\nu$ in \cm, of CH$_4\cdot$Ar,
 computed with the GENIUSH--Smolyak approach \cite{AvMa19a}
 and the \pesname\ PES developed in the present work.
 \label{tab:vib}
}
\scalebox{0.9}{%
\begin{tabular}{@{}c@{}cc cccc c@{\ \ }  cc@{\ }cc  @{\ \ } c@{\ }cc  @{\ \ } c@{\ }cc @{}}
\cline{1-17}\\[-0.4cm]
\cline{1-17}\\[-0.4cm]
&~~&
\multicolumn{3}{c}{Assignment$^\text{a}$} &&
\multicolumn{5}{c}{12D$^\text{b}$} &~~& 
\multicolumn{2}{c}{3D($\langle r\rangle_0$)$^\text{c}$} &~~&
\multicolumn{2}{c}{3D($\langle B\rangle_0$)$^\text{d}$} \\
\cline{3-5}\cline{7-11}\cline{13-14}\cline{16-17} \\[-0.35cm]
\#	&& $j$ & $n_R$ & $\Gamma$ & &
$\Delta_0$&
$\Delta_1$&
$\Delta_2$&
$\tilde\nu(b=3)$& 
&& 
$\tilde\nu_\text{3D}$	& $\delta$$^\text{e}$ && 
$\tilde\nu_\text{3D}$	& $\delta$$^\text{e}$ \\
\cline{1-17}\\[-0.4cm]
0$^\text{f}$ &&	0& 0	&	A$_1$	&&66.10	   &   40.67 & 38.32	&	9745.40	&	&&      54.30	&	--	&&      54.38	&	--	\\[0.2cm]
1--3	&&	1& 0	&	F$_2$	&& 0.21	   &    0.03 & 0.00	&	8.68	&	&&	8.65	&	0.04	&&	8.87	&	$-$0.19	\\
4	&&	0& 1	&	A$_1$	&& 0.10	   &    0.01 & 0.00	&	29.88	&	&&	29.80	&	0.07	&&	29.89	&	$-$0.01	\\
5--7	&&	2& 0	&	F$_2$	&& 0.29	   &    0.01 & $-$0.01	&	31.63	&	&&	30.84	&	0.79	&&	31.29	&	0.35	\\
8--9	&&	2& 0	&	E	&& 0.40	   &    0.05 & 0.01	&	32.34	&	&&	31.65	&	0.69	&&	32.21	&	0.13	\\
10--12	&&	1& 1	&	F$_2$	&& 0.21	   &    0.02 & 0.01	&	45.76	&	&&	45.27	&	0.49	&&	45.63	&	0.13	\\
13	&&	0& 2	&	A$_1$	&& 0.06	   &    0.00 & 0.01	&	54.79	&	&&	54.54	&	0.26	&&	54.66	&	0.14	\\
14--16	&&	2& 1	&	F$_2$	&& 0.50	   &    0.05 & 0.00	&	56.80	&	&&	55.92	&	0.88	&&	56.58	&	0.22	\\
17--18	&&	2& 1	&	E	&& 0.39	   &    0.05 & 0.02	&	65.64	&	&&	64.92	&	0.72	&&	65.49	&	0.14	\\
19--21	&&	3& 0	&	F$_2$	&& 0.80	   &    0.08 & $-$0.01	&	66.08	&	&&	64.52	&	1.56	&&	65.72	&	0.36	\\
22--24	&&	1& 2	&	F$_1$	&& 0.33	   &    0.03 & 0.01	&	67.86	&	&&	66.86	&	1.00	&&	67.45	&	0.41	\\
25	&&	0& 3	&	A$_1$	&& 0.30	   &    0.02 & 0.00	&	72.62	&	&&	71.19	&	1.42	&&	72.02	&	0.60	\\
26	&&	3& 0	&	A$_1$	&& 0.31	   &    0.02 & 0.01	&	75.23	&	&&	74.63	&	0.60	&&	74.97	&	0.26	\\
27--29	&&	2& 2	&	F$_2$	&& 0.29	   &    0.02 & 0.00	&	77.73	&	&&	76.46	&	1.28	&&	77.11	&	0.63	\\
30--32	&&	1& 3	&	F$_2$	&& 0.32	   &    0.03 & 0.01	&	82.25	&	&&	81.71	&	0.54	&&	82.17	&	0.08	\\
33	&&	0& 4	&	A$_1$	&& $-$0.12 & $-$0.01 & 0.02	&	87.53	&	&&	87.06	&	0.47	&&	87.18	&	0.35	\\
34--36	&&	1& 4	&	F$_2$	&& 0.23	   &    0.03 & 0.02	&	91.42	&	&&	90.57	&	0.84	&&	91.07	&	0.35	\\
37--38	&&	2& 4	&	E	&& 0.34	   &    0.04 & 0.02	&	91.61	&	&&	90.80	&	0.81	&&	91.38	&	0.23	\\
39	&&	0& 5	&	A$_1$	&& $-$0.42 & $-$0.26 & 0.03	&	95.44	&	&&	94.75	&	0.69	&&	95.02	&	0.42	\\
40--42	&&	3& 3	&       F$_2$	&& 0.97	   &    0.10 & 0.00	&	95.69	&	&&	94.07	&	1.61	&&	95.42	&	0.26	\\
43	&&	0& 6	&	A$_1$	&& $-$0.09 &    0.03 & 0.03	&	99.11	&	&&	98.21	&	0.90	&&	98.72	&	0.39	\\
44--46	&&	 &	&		&& --	   & --    & --          &	--      &	&&	99.32	&	--	&&	--	&	--	\\
\cline{1-17}\\[-0.4cm]
rms$^\text{g}$	&&		&		&& & 0.40 & 0.07  &	0.01	&		&&		&&	0.93	&&		&	0.32	\\
\cline{1-17}\\[-0.4cm]
\cline{1-17}\\[-0.4cm]
\end{tabular}
}
\begin{flushleft}
$^\text{a}$
  Characterization of the computed states using the 3D wave functions of the 3D(fit) column. 
  $j$: angular momentum quantum number of the methane fragment and the relative diatom
  in the $[j,j]_{00}$ dominant coupled-rotor (CR) function \cite{FeMa19};
  $n_R$: vibrational excitation along $R$;
  $\Gamma$: irrep label of the $T_\text{d}$(M) molecular symmetry group of the complex 
  based on the CR assignment and irrep decomposition \cite{FeMa19}. \\
$^\text{b}$
  $\Delta_k=\tilde\nu(b=k)-\tilde\nu(b=3)$, where $b$ is the pruning parameter of the methane basis
  the intermolecular radial and angular representations are defined in the text and is 
  sufficient for converging the figures shown in the table. \\
$^\text{c}$
  3D computation using rigorous geometrical constraints with
  a regular tetrahedral methane structure with 
  $\langle r_{\text{C--H}} \rangle_0=2.093\,624\,127$~bohr
  (used both in the KEO and in the PES). \\
$^\text{d}$
  3D computation using an `adjusted'
  $r_\text{fit}(\text{C--H})=2.072\,988\,169$~bohr %2.072988169084743 
  C--H distance in the KEO, which in a 
  2D coupled-rotor computation \cite{SaCsMa17,FeMa19}
  reproduces the $\langle B\rangle_0=5.212\ 508\ 664$~\cm\ 
  effective rotational constant corresponding to this PES.  
  To define the 3D cut of the PES,
  we used $\langle r_\text{C--H} \rangle_0$. \\
%   {\colred ??? $B_\text{fit}=5.212507$ cm$^{-1}$} \\
%
$^\text{e}$
  $\delta=\tilde\nu(b=3)-\tilde\nu_\text{3D}$. \\
$^\text{f}$
  Zero-point vibration of the complex. The vibrational excitation energies, \#: 1,2,3,\ldots, 
  are given with respect to this value. \\

$^\text{g}$
  Root-mean-square deviation from the $\tilde\nu(b=3)$ (12D) result. \\
\end{flushleft}
\end{table}
}

\clearpage                    
\section{Summary and conclusions}
\noindent %
The present work reports the development of a full-dimensional, near-spectroscopic quality
\emph{ab initio} potential energy surface for the van-der-Waals complex of the methane molecule and 
an argon atom.
The PES development is accompanied with the computation of all vibrational bound states of 
this complex including all (12) vibrational degrees of freedom 
in a near-variational treatment using the GENIUSH program and the Smolyak algorithm.
The vibrational excitation energies obtained within a 12D treatment
were used to assess traditional 3D (rigid-monomer) approaches.
With further development of the quantum dynamics methodology, 
full-dimensional computations will 
become more widespread and applicable to floppy molecules
or molecular complexes over a broad energy range.

%%%%%%%%%%%%%%%%%%%%%%%%%%%%%%%%%%%%%%%%%%%%%%%%%%%%%%%%%%%%%%%%%%%%%%%%%%
% Acknowledgment
%%%%%%%%%%%%%%%%%%%%%%%%%%%%%%%%%%%%%%%%%%%%%%%%%%%%%%%%%%
\vspace{1cm}
\noindent %
\paragraph*{Conflicts of interest} 
There are no conflicts of interest to declare.

\vspace{0.5cm}
\paragraph*{Acknowledgement}
\noindent %
We thank Xiao-Gang Wang and Tucker Carrington for sharing their well-tested
sincot-Legendre-DVR implementation (codvr.f90) with us.
G.A. and E.M. acknowledge financial support from a PROMYS Grant (no. IZ11Z0\_166525)  
of the Swiss National Science Foundation. 
D.P. and G.C. thank the National Research, Development and Innovation Office-NKFIH, K-125317, 
the Ministry of Human Capacities, Hungary grant 20391-3/2018/FEKUSTRAT, and 
the Momentum (Lend\"ulet) Program of the Hungarian Academy of Sciences for financial support.

\vspace{0.5cm}
\paragraph*{Electronic supplementary information (ESI) available: }
Coefficients of the FullD-2019 PES and normal coordinate coefficients used for the vibrational
computations. For additional information about the PES libraries please contact G.C.

% \bibliography{paper.bib} 

\begin{thebibliography}{61}%
\makeatletter
\providecommand \@ifxundefined [1]{%
 \@ifx{#1\undefined}
}%
\providecommand \@ifnum [1]{%
 \ifnum #1\expandafter \@firstoftwo
 \else \expandafter \@secondoftwo
 \fi
}%
\providecommand \@ifx [1]{%
 \ifx #1\expandafter \@firstoftwo
 \else \expandafter \@secondoftwo
 \fi
}%
\providecommand \natexlab [1]{#1}%
\providecommand \enquote  [1]{``#1''}%
\providecommand \bibnamefont  [1]{#1}%
\providecommand \bibfnamefont [1]{#1}%
\providecommand \citenamefont [1]{#1}%
\providecommand \href@noop [0]{\@secondoftwo}%
\providecommand \href [0]{\begingroup \@sanitize@url \@href}%
\providecommand \@href[1]{\@@startlink{#1}\@@href}%
\providecommand \@@href[1]{\endgroup#1\@@endlink}%
\providecommand \@sanitize@url [0]{\catcode `\\12\catcode `\$12\catcode
  `\&12\catcode `\#12\catcode `\^12\catcode `\_12\catcode `\%12\relax}%
\providecommand \@@startlink[1]{}%
\providecommand \@@endlink[0]{}%
\providecommand \url  [0]{\begingroup\@sanitize@url \@url }%
\providecommand \@url [1]{\endgroup\@href {#1}{\urlprefix }}%
\providecommand \urlprefix  [0]{URL }%
\providecommand \Eprint [0]{\href }%
\providecommand \doibase [0]{http://dx.doi.org/}%
\providecommand \selectlanguage [0]{\@gobble}%
\providecommand \bibinfo  [0]{\@secondoftwo}%
\providecommand \bibfield  [0]{\@secondoftwo}%
\providecommand \translation [1]{[#1]}%
\providecommand \BibitemOpen [0]{}%
\providecommand \bibitemStop [0]{}%
\providecommand \bibitemNoStop [0]{.\EOS\space}%
\providecommand \EOS [0]{\spacefactor3000\relax}%
\providecommand \BibitemShut  [1]{\csname bibitem#1\endcsname}%
\let\auto@bib@innerbib\@empty
%</preamble>
\bibitem [{\citenamefont {Murrell}\ \emph {et~al.}(1984)\citenamefont
  {Murrell}, \citenamefont {Carter}, \citenamefont {Farantos}, \citenamefont
  {Huxley},\ and\ \citenamefont {Varandas}}]{MuCaFaHuVa84}%
  \BibitemOpen
  \bibfield  {author} {\bibinfo {author} {\bibfnamefont {J.~N.}\ \bibnamefont
  {Murrell}}, \bibinfo {author} {\bibfnamefont {S.}~\bibnamefont {Carter}},
  \bibinfo {author} {\bibfnamefont {S.~C.}\ \bibnamefont {Farantos}}, \bibinfo
  {author} {\bibfnamefont {P.}~\bibnamefont {Huxley}}, \ and\ \bibinfo {author}
  {\bibfnamefont {A.~J.~C.}\ \bibnamefont {Varandas}},\ }\href@noop {} {\emph
  {\bibinfo {title} {Molecular Potential Energy Functions}}}\ (\bibinfo
  {publisher} {Wiley},\ \bibinfo {address} {New York},\ \bibinfo {year}
  {1984})\BibitemShut {NoStop}%
\bibitem [{\citenamefont {Quack}\ and\ \citenamefont {Suhm}(1991)}]{QuSu91}%
  \BibitemOpen
  \bibfield  {author} {\bibinfo {author} {\bibfnamefont {M.}~\bibnamefont
  {Quack}}\ and\ \bibinfo {author} {\bibfnamefont {M.~A.}\ \bibnamefont
  {Suhm}},\ }\href@noop {} {\bibfield  {journal} {\bibinfo  {journal} {J. Chem.
  Phys.}\ }\textbf {\bibinfo {volume} {95}},\ \bibinfo {pages} {28} (\bibinfo
  {year} {1991})}\BibitemShut {NoStop}%
\bibitem [{\citenamefont {Quack}\ \emph {et~al.}(1993)\citenamefont {Quack},
  \citenamefont {Stohner},\ and\ \citenamefont {Suhm}}]{QuStSu93}%
  \BibitemOpen
  \bibfield  {author} {\bibinfo {author} {\bibfnamefont {M.}~\bibnamefont
  {Quack}}, \bibinfo {author} {\bibfnamefont {J.}~\bibnamefont {Stohner}}, \
  and\ \bibinfo {author} {\bibfnamefont {M.~A.}\ \bibnamefont {Suhm}},\
  }\href@noop {} {\bibfield  {journal} {\bibinfo  {journal} {J. Mol. Struct.}\
  }\textbf {\bibinfo {volume} {294}},\ \bibinfo {pages} {33} (\bibinfo {year}
  {1993})}\BibitemShut {NoStop}%
\bibitem [{\citenamefont {Groenenboom}\ \emph {et~al.}(2000)\citenamefont
  {Groenenboom}, \citenamefont {Mas}, \citenamefont {Bukowski}, \citenamefont
  {Szalewicz}, \citenamefont {Wormer},\ and\ \citenamefont {van~der
  Avoird}}]{Groenenboom:00}%
  \BibitemOpen
  \bibfield  {author} {\bibinfo {author} {\bibfnamefont {G.~C.}\ \bibnamefont
  {Groenenboom}}, \bibinfo {author} {\bibfnamefont {E.~M.}\ \bibnamefont
  {Mas}}, \bibinfo {author} {\bibfnamefont {R.}~\bibnamefont {Bukowski}},
  \bibinfo {author} {\bibfnamefont {K.}~\bibnamefont {Szalewicz}}, \bibinfo
  {author} {\bibfnamefont {P.~E.~S.}\ \bibnamefont {Wormer}}, \ and\ \bibinfo
  {author} {\bibfnamefont {A.}~\bibnamefont {van~der Avoird}},\ }\href@noop {}
  {\bibfield  {journal} {\bibinfo  {journal} {Phys. Rev. Lett.}\ }\textbf
  {\bibinfo {volume} {84}},\ \bibinfo {pages} {4072} (\bibinfo {year}
  {2000})}\BibitemShut {NoStop}%
\bibitem [{\citenamefont {Cencek}\ \emph {et~al.}(2008)\citenamefont {Cencek},
  \citenamefont {Szalewicz}, \citenamefont {Leforestier}, \citenamefont {van
  Harrevelt},\ and\ \citenamefont {van~der Avoird}}]{Cencek:08b}%
  \BibitemOpen
  \bibfield  {author} {\bibinfo {author} {\bibfnamefont {W.}~\bibnamefont
  {Cencek}}, \bibinfo {author} {\bibfnamefont {K.}~\bibnamefont {Szalewicz}},
  \bibinfo {author} {\bibfnamefont {C.}~\bibnamefont {Leforestier}}, \bibinfo
  {author} {\bibfnamefont {R.}~\bibnamefont {van Harrevelt}}, \ and\ \bibinfo
  {author} {\bibfnamefont {A.}~\bibnamefont {van~der Avoird}},\ }\href@noop {}
  {\bibfield  {journal} {\bibinfo  {journal} {Phys. Chem. Chem. Phys.}\
  }\textbf {\bibinfo {volume} {10}},\ \bibinfo {pages} {4716} (\bibinfo {year}
  {2008})}\BibitemShut {NoStop}%
\bibitem [{\citenamefont {Wang}\ and\ \citenamefont {Bowman}(2011)}]{WaBo11}%
  \BibitemOpen
  \bibfield  {author} {\bibinfo {author} {\bibfnamefont {Y.}~\bibnamefont
  {Wang}}\ and\ \bibinfo {author} {\bibfnamefont {J.~M.}\ \bibnamefont
  {Bowman}},\ }\href@noop {} {\bibfield  {journal} {\bibinfo  {journal} {J.
  Chem. Phys.}\ }\textbf {\bibinfo {volume} {134}},\ \bibinfo {pages} {154510}
  (\bibinfo {year} {2011})}\BibitemShut {NoStop}%
\bibitem [{\citenamefont {Medders}\ \emph {et~al.}(2013)\citenamefont
  {Medders}, \citenamefont {Babin},\ and\ \citenamefont {Paesani}}]{MeBaPa13}%
  \BibitemOpen
  \bibfield  {author} {\bibinfo {author} {\bibfnamefont {G.~R.}\ \bibnamefont
  {Medders}}, \bibinfo {author} {\bibfnamefont {V.}~\bibnamefont {Babin}}, \
  and\ \bibinfo {author} {\bibfnamefont {F.}~\bibnamefont {Paesani}},\
  }\href@noop {} {\bibfield  {journal} {\bibinfo  {journal} {J. Chem. Theory
  Comput.}\ }\textbf {\bibinfo {volume} {9}},\ \bibinfo {pages} {1103}
  (\bibinfo {year} {2013})}\BibitemShut {NoStop}%
\bibitem [{\citenamefont {Babin}\ \emph {et~al.}(2014)\citenamefont {Babin},
  \citenamefont {Medders},\ and\ \citenamefont {Paesani}}]{BaMePa14}%
  \BibitemOpen
  \bibfield  {author} {\bibinfo {author} {\bibfnamefont {V.}~\bibnamefont
  {Babin}}, \bibinfo {author} {\bibfnamefont {G.~R.}\ \bibnamefont {Medders}},
  \ and\ \bibinfo {author} {\bibfnamefont {F.}~\bibnamefont {Paesani}},\
  }\href@noop {} {\bibfield  {journal} {\bibinfo  {journal} {J. Chem. Theory
  Comput.}\ }\textbf {\bibinfo {volume} {10}},\ \bibinfo {pages} {1599}
  (\bibinfo {year} {2014})}\BibitemShut {NoStop}%
\bibitem [{\citenamefont {Jankowski}\ \emph {et~al.}(2015)\citenamefont
  {Jankowski}, \citenamefont {Murdachaew}, \citenamefont {Bukowski},
  \citenamefont {Akin-Ojo}, \citenamefont {Leforestier},\ and\ \citenamefont
  {Szalewicz}}]{Jankowski:15}%
  \BibitemOpen
  \bibfield  {author} {\bibinfo {author} {\bibfnamefont {P.}~\bibnamefont
  {Jankowski}}, \bibinfo {author} {\bibfnamefont {G.}~\bibnamefont
  {Murdachaew}}, \bibinfo {author} {\bibfnamefont {R.}~\bibnamefont
  {Bukowski}}, \bibinfo {author} {\bibfnamefont {O.}~\bibnamefont {Akin-Ojo}},
  \bibinfo {author} {\bibfnamefont {C.}~\bibnamefont {Leforestier}}, \ and\
  \bibinfo {author} {\bibfnamefont {K.}~\bibnamefont {Szalewicz}},\ }\href@noop
  {} {\bibfield  {journal} {\bibinfo  {journal} {J. Phys. Chem. A}\ }\textbf
  {\bibinfo {volume} {119}},\ \bibinfo {pages} {2940} (\bibinfo {year}
  {2015})}\BibitemShut {NoStop}%
\bibitem [{\citenamefont {G{\'o}ra}\ \emph {et~al.}(2014)\citenamefont
  {G{\'o}ra}, \citenamefont {Cencek}, \citenamefont {Podeszwa}, \citenamefont
  {van~der Avoird},\ and\ \citenamefont {Szalewicz}}]{Gora:14}%
  \BibitemOpen
  \bibfield  {author} {\bibinfo {author} {\bibfnamefont {U.}~\bibnamefont
  {G{\'o}ra}}, \bibinfo {author} {\bibfnamefont {W.}~\bibnamefont {Cencek}},
  \bibinfo {author} {\bibfnamefont {R.}~\bibnamefont {Podeszwa}}, \bibinfo
  {author} {\bibfnamefont {A.}~\bibnamefont {van~der Avoird}}, \ and\ \bibinfo
  {author} {\bibfnamefont {K.}~\bibnamefont {Szalewicz}},\ }\href@noop {}
  {\bibfield  {journal} {\bibinfo  {journal} {J. Chem. Phys.}\ }\textbf
  {\bibinfo {volume} {140}},\ \bibinfo {pages} {194101} (\bibinfo {year}
  {2014})}\BibitemShut {NoStop}%
\bibitem [{\citenamefont {Heijmen}\ \emph {et~al.}(1999)\citenamefont
  {Heijmen}, \citenamefont {Wormer}, \citenamefont {van~der Avoird},
  \citenamefont {Miller},\ and\ \citenamefont {Moszynski}}]{HeWoAvMiMo99}%
  \BibitemOpen
  \bibfield  {author} {\bibinfo {author} {\bibfnamefont {T.~G.~A.}\
  \bibnamefont {Heijmen}}, \bibinfo {author} {\bibfnamefont {P.~E.~S.}\
  \bibnamefont {Wormer}}, \bibinfo {author} {\bibfnamefont {A.}~\bibnamefont
  {van~der Avoird}}, \bibinfo {author} {\bibfnamefont {R.~E.}\ \bibnamefont
  {Miller}}, \ and\ \bibinfo {author} {\bibfnamefont {R.}~\bibnamefont
  {Moszynski}},\ }\href@noop {} {\bibfield  {journal} {\bibinfo  {journal} {J.
  Chem. Phys.}\ }\textbf {\bibinfo {volume} {110}},\ \bibinfo {pages} {5639}
  (\bibinfo {year} {1999})}\BibitemShut {NoStop}%
\bibitem [{\citenamefont {Sarka}\ \emph {et~al.}(2016)\citenamefont {Sarka},
  \citenamefont {Cs\'asz\'ar}, \citenamefont {Althorpe}, \citenamefont
  {Wales},\ and\ \citenamefont {M\'atyus}}]{SaCsAlWaMa16}%
  \BibitemOpen
  \bibfield  {author} {\bibinfo {author} {\bibfnamefont {J.}~\bibnamefont
  {Sarka}}, \bibinfo {author} {\bibfnamefont {A.~G.}\ \bibnamefont
  {Cs\'asz\'ar}}, \bibinfo {author} {\bibfnamefont {S.~C.}\ \bibnamefont
  {Althorpe}}, \bibinfo {author} {\bibfnamefont {D.~J.}\ \bibnamefont {Wales}},
  \ and\ \bibinfo {author} {\bibfnamefont {E.}~\bibnamefont {M\'atyus}},\
  }\href@noop {} {\bibfield  {journal} {\bibinfo  {journal} {Phys. Chem. Chem.
  Phys.}\ }\textbf {\bibinfo {volume} {18}},\ \bibinfo {pages} {22816}
  (\bibinfo {year} {2016})}\BibitemShut {NoStop}%
\bibitem [{\citenamefont {Sarka}\ \emph {et~al.}(2017)\citenamefont {Sarka},
  \citenamefont {Cs\'asz\'ar},\ and\ \citenamefont {M\'atyus}}]{SaCsMa17}%
  \BibitemOpen
  \bibfield  {author} {\bibinfo {author} {\bibfnamefont {J.}~\bibnamefont
  {Sarka}}, \bibinfo {author} {\bibfnamefont {A.~G.}\ \bibnamefont
  {Cs\'asz\'ar}}, \ and\ \bibinfo {author} {\bibfnamefont {E.}~\bibnamefont
  {M\'atyus}},\ }\href@noop {} {\bibfield  {journal} {\bibinfo  {journal}
  {Phys. Chem. Chem. Phys.}\ }\textbf {\bibinfo {volume} {2}},\ \bibinfo
  {pages} {15335} (\bibinfo {year} {2017})}\BibitemShut {NoStop}%
\bibitem [{\citenamefont {Metz}\ \emph {et~al.}(2019)\citenamefont {Metz},
  \citenamefont {Szalewicz}, \citenamefont {Sarka}, \citenamefont {T\'obi\'as},
  \citenamefont {Cs\'asz\'ar},\ and\ \citenamefont
  {M\'atyus}}]{MeSzSaToCsMa19}%
  \BibitemOpen
  \bibfield  {author} {\bibinfo {author} {\bibfnamefont {M.~P.}\ \bibnamefont
  {Metz}}, \bibinfo {author} {\bibfnamefont {K.}~\bibnamefont {Szalewicz}},
  \bibinfo {author} {\bibfnamefont {J.}~\bibnamefont {Sarka}}, \bibinfo
  {author} {\bibfnamefont {R.}~\bibnamefont {T\'obi\'as}}, \bibinfo {author}
  {\bibfnamefont {A.~G.}\ \bibnamefont {Cs\'asz\'ar}}, \ and\ \bibinfo {author}
  {\bibfnamefont {E.}~\bibnamefont {M\'atyus}},\ }\href@noop {} {\bibfield
  {journal} {\bibinfo  {journal} {Phys. Chem. Chem. Phys.}\ }\textbf {\bibinfo
  {volume} {21}},\ \bibinfo {pages} {13504} (\bibinfo {year}
  {2019})}\BibitemShut {NoStop}%
\bibitem [{\citenamefont {Jeziorska}\ \emph {et~al.}(2000)\citenamefont
  {Jeziorska}, \citenamefont {Jankowski}, \citenamefont {Szalewicz},\ and\
  \citenamefont {Jeziorski}}]{JeJaSzJe00}%
  \BibitemOpen
  \bibfield  {author} {\bibinfo {author} {\bibfnamefont {M.}~\bibnamefont
  {Jeziorska}}, \bibinfo {author} {\bibfnamefont {P.}~\bibnamefont
  {Jankowski}}, \bibinfo {author} {\bibfnamefont {K.}~\bibnamefont
  {Szalewicz}}, \ and\ \bibinfo {author} {\bibfnamefont {B.}~\bibnamefont
  {Jeziorski}},\ }\href@noop {} {\bibfield  {journal} {\bibinfo  {journal} {J.
  Chem. Phys.}\ }\textbf {\bibinfo {volume} {113}},\ \bibinfo {pages} {2957}
  (\bibinfo {year} {2000})}\BibitemShut {NoStop}%
\bibitem [{\citenamefont {Wang}\ and\ \citenamefont {{Carrington,
  Jr.}}(2017)}]{WaCa17}%
  \BibitemOpen
  \bibfield  {author} {\bibinfo {author} {\bibfnamefont {X.-G.}\ \bibnamefont
  {Wang}}\ and\ \bibinfo {author} {\bibfnamefont {T.}~\bibnamefont
  {{Carrington, Jr.}}},\ }\href@noop {} {\bibfield  {journal} {\bibinfo
  {journal} {J. Chem. Phys.}\ }\textbf {\bibinfo {volume} {146}},\ \bibinfo
  {pages} {104105} (\bibinfo {year} {2017})}\BibitemShut {NoStop}%
\bibitem [{\citenamefont {Wang}\ and\ \citenamefont {{Carrington,
  Jr.}}(2018)}]{WaCa18WW}%
  \BibitemOpen
  \bibfield  {author} {\bibinfo {author} {\bibfnamefont {X.-G.}\ \bibnamefont
  {Wang}}\ and\ \bibinfo {author} {\bibfnamefont {T.}~\bibnamefont
  {{Carrington, Jr.}}},\ }\href@noop {} {\bibfield  {journal} {\bibinfo
  {journal} {J. Chem. Phys.}\ }\textbf {\bibinfo {volume} {148}},\ \bibinfo
  {pages} {074108} (\bibinfo {year} {2018})}\BibitemShut {NoStop}%
\bibitem [{\citenamefont {Avila}\ and\ \citenamefont {Matyus}(2019)}]{AvMa19b}%
  \BibitemOpen
  \bibfield  {author} {\bibinfo {author} {\bibfnamefont {G.}~\bibnamefont
  {Avila}}\ and\ \bibinfo {author} {\bibfnamefont {E.}~\bibnamefont {Matyus}},\
  }\href@noop {} {\bibfield  {journal} {\bibinfo  {journal} {J. Chem. Phys.}\
  }\textbf {\bibinfo {volume} {151}},\ \bibinfo {pages} {154301} (\bibinfo
  {year} {2019})}\BibitemShut {NoStop}%
\bibitem [{\citenamefont {F\'abri}\ \emph {et~al.}(2013)\citenamefont
  {F\'abri}, \citenamefont {Cs\'asz\'ar},\ and\ \citenamefont
  {Czak\'o}}]{FaCsCz13}%
  \BibitemOpen
  \bibfield  {author} {\bibinfo {author} {\bibfnamefont {C.}~\bibnamefont
  {F\'abri}}, \bibinfo {author} {\bibfnamefont {A.~G.}\ \bibnamefont
  {Cs\'asz\'ar}}, \ and\ \bibinfo {author} {\bibfnamefont {G.}~\bibnamefont
  {Czak\'o}},\ }\href@noop {} {\bibfield  {journal} {\bibinfo  {journal} {J.
  Phys. Chem. A}\ }\textbf {\bibinfo {volume} {117}},\ \bibinfo {pages} {6975}
  (\bibinfo {year} {2013})}\BibitemShut {NoStop}%
\bibitem [{\citenamefont {Wodraszka}\ \emph {et~al.}(2012)\citenamefont
  {Wodraszka}, \citenamefont {Palma},\ and\ \citenamefont {Manthe}}]{WoPaMa12}%
  \BibitemOpen
  \bibfield  {author} {\bibinfo {author} {\bibfnamefont {R.}~\bibnamefont
  {Wodraszka}}, \bibinfo {author} {\bibfnamefont {J.}~\bibnamefont {Palma}}, \
  and\ \bibinfo {author} {\bibfnamefont {U.}~\bibnamefont {Manthe}},\
  }\href@noop {} {\bibfield  {journal} {\bibinfo  {journal} {J. Phys. Chem. A}\
  }\textbf {\bibinfo {volume} {116}},\ \bibinfo {pages} {11249} (\bibinfo
  {year} {2012})}\BibitemShut {NoStop}%
\bibitem [{\citenamefont {Papp}\ \emph {et~al.}(2017)\citenamefont {Papp},
  \citenamefont {Sarka}, \citenamefont {Szidarovszky}, \citenamefont
  {Cs\'asz\'ar.}, \citenamefont {M\'atyus}, \citenamefont {Hochlaf},\ and\
  \citenamefont {Stoecklin}}]{ArNOp}%
  \BibitemOpen
  \bibfield  {author} {\bibinfo {author} {\bibfnamefont {D.}~\bibnamefont
  {Papp}}, \bibinfo {author} {\bibfnamefont {J.}~\bibnamefont {Sarka}},
  \bibinfo {author} {\bibfnamefont {T.}~\bibnamefont {Szidarovszky}}, \bibinfo
  {author} {\bibfnamefont {A.~G.}\ \bibnamefont {Cs\'asz\'ar.}}, \bibinfo
  {author} {\bibfnamefont {E.}~\bibnamefont {M\'atyus}}, \bibinfo {author}
  {\bibfnamefont {M.}~\bibnamefont {Hochlaf}}, \ and\ \bibinfo {author}
  {\bibfnamefont {T.}~\bibnamefont {Stoecklin}},\ }\href {\doibase
  10.1039/C6CP07731E} {\bibfield  {journal} {\bibinfo  {journal} {Phys. Chem.
  Chem. Phys.}\ }\textbf {\bibinfo {volume} {19}},\ \bibinfo {pages} {8152}
  (\bibinfo {year} {2017})}\BibitemShut {NoStop}%
\bibitem [{\citenamefont {Felker}\ and\ \citenamefont {Bacic}(2019)}]{FeBa19}%
  \BibitemOpen
  \bibfield  {author} {\bibinfo {author} {\bibfnamefont {P.~M.}\ \bibnamefont
  {Felker}}\ and\ \bibinfo {author} {\bibfnamefont {Z.}~\bibnamefont {Bacic}},\
  }\href@noop {} {\bibfield  {journal} {\bibinfo  {journal} {J. Chem. Phys.}\
  }\textbf {\bibinfo {volume} {151}},\ \bibinfo {pages} {024305} (\bibinfo
  {year} {2019})}\BibitemShut {NoStop}%
\bibitem [{\citenamefont {Bowman}\ \emph {et~al.}(2003)\citenamefont {Bowman},
  \citenamefont {Carter},\ and\ \citenamefont {Huang}}]{MM2}%
  \BibitemOpen
  \bibfield  {author} {\bibinfo {author} {\bibfnamefont {J.~M.}\ \bibnamefont
  {Bowman}}, \bibinfo {author} {\bibfnamefont {S.}~\bibnamefont {Carter}}, \
  and\ \bibinfo {author} {\bibfnamefont {X.}~\bibnamefont {Huang}},\
  }\href@noop {} {\bibfield  {journal} {\bibinfo  {journal} {International
  Reviews in Physical Chemistry}\ }\textbf {\bibinfo {volume} {22}},\ \bibinfo
  {pages} {533} (\bibinfo {year} {2003})}\BibitemShut {NoStop}%
\bibitem [{\citenamefont {Avila}\ and\ \citenamefont
  {T.~Carrington}(2009)}]{tc-gab1}%
  \BibitemOpen
  \bibfield  {author} {\bibinfo {author} {\bibfnamefont {G.}~\bibnamefont
  {Avila}}\ and\ \bibinfo {author} {\bibfnamefont {J.}~\bibnamefont
  {T.~Carrington}},\ }\href@noop {} {\bibfield  {journal} {\bibinfo  {journal}
  {J. Chem. Phys.}\ }\textbf {\bibinfo {volume} {131}},\ \bibinfo {pages}
  {174103} (\bibinfo {year} {2009})}\BibitemShut {NoStop}%
\bibitem [{\citenamefont {Avila}\ and\ \citenamefont
  {T.~Carrington}(2011{\natexlab{a}})}]{tc-gab2}%
  \BibitemOpen
  \bibfield  {author} {\bibinfo {author} {\bibfnamefont {G.}~\bibnamefont
  {Avila}}\ and\ \bibinfo {author} {\bibfnamefont {J.}~\bibnamefont
  {T.~Carrington}},\ }\href@noop {} {\bibfield  {journal} {\bibinfo  {journal}
  {J. Chem. Phys.}\ }\textbf {\bibinfo {volume} {134}},\ \bibinfo {pages}
  {054126} (\bibinfo {year} {2011}{\natexlab{a}})}\BibitemShut {NoStop}%
\bibitem [{\citenamefont {Avila}\ and\ \citenamefont
  {T.~Carrington}(2011{\natexlab{b}})}]{AvCa11b}%
  \BibitemOpen
  \bibfield  {author} {\bibinfo {author} {\bibfnamefont {G.}~\bibnamefont
  {Avila}}\ and\ \bibinfo {author} {\bibfnamefont {J.}~\bibnamefont
  {T.~Carrington}},\ }\href@noop {} {\bibfield  {journal} {\bibinfo  {journal}
  {J. Chem. Phys.}\ }\textbf {\bibinfo {volume} {134}},\ \bibinfo {pages}
  {064101} (\bibinfo {year} {2011}{\natexlab{b}})}\BibitemShut {NoStop}%
\bibitem [{\citenamefont {Leclerc}\ and\ \citenamefont
  {Carrington}(2014)}]{CP1}%
  \BibitemOpen
  \bibfield  {author} {\bibinfo {author} {\bibfnamefont {A.}~\bibnamefont
  {Leclerc}}\ and\ \bibinfo {author} {\bibfnamefont {T.}~\bibnamefont
  {Carrington}},\ }\href@noop {} {\bibfield  {journal} {\bibinfo  {journal} {J.
  Chem. Phys.}\ }\textbf {\bibinfo {volume} {140}},\ \bibinfo {pages} {174111}
  (\bibinfo {year} {2014})}\BibitemShut {NoStop}%
\bibitem [{\citenamefont {Thomas}\ and\ \citenamefont
  {Carrington}(2017)}]{CP2}%
  \BibitemOpen
  \bibfield  {author} {\bibinfo {author} {\bibfnamefont {P.~S.}\ \bibnamefont
  {Thomas}}\ and\ \bibinfo {author} {\bibfnamefont {T.}~\bibnamefont
  {Carrington}},\ }\href@noop {} {\bibfield  {journal} {\bibinfo  {journal} {J.
  Chem. Phys.}\ }\textbf {\bibinfo {volume} {146}},\ \bibinfo {pages} {204110}
  (\bibinfo {year} {2017})}\BibitemShut {NoStop}%
\bibitem [{\citenamefont {Halverson}\ and\ \citenamefont
  {Poirier}(2015)}]{betterpr1}%
  \BibitemOpen
  \bibfield  {author} {\bibinfo {author} {\bibfnamefont {T.}~\bibnamefont
  {Halverson}}\ and\ \bibinfo {author} {\bibfnamefont {B.}~\bibnamefont
  {Poirier}},\ }\href@noop {} {\bibfield  {journal} {\bibinfo  {journal} {The
  Journal of Physical Chemistry A}\ }\textbf {\bibinfo {volume} {119}},\
  \bibinfo {pages} {12417} (\bibinfo {year} {2015})}\BibitemShut {NoStop}%
\bibitem [{\citenamefont {Brown}\ and\ \citenamefont
  {Carrington}(2016)}]{betterpr2}%
  \BibitemOpen
  \bibfield  {author} {\bibinfo {author} {\bibfnamefont {J.}~\bibnamefont
  {Brown}}\ and\ \bibinfo {author} {\bibfnamefont {T.}~\bibnamefont
  {Carrington}},\ }\href@noop {} {\bibfield  {journal} {\bibinfo  {journal} {J.
  Chem. Phys.}\ }\textbf {\bibinfo {volume} {145}},\ \bibinfo {pages} {144104}
  (\bibinfo {year} {2016})}\BibitemShut {NoStop}%
\bibitem [{\citenamefont {Madsen}\ \emph {et~al.}(2018)\citenamefont {Madsen},
  \citenamefont {Godtliebsen}, \citenamefont {Losilla},\ and\ \citenamefont
  {Christiansen}}]{MaGoLoCh18}%
  \BibitemOpen
  \bibfield  {author} {\bibinfo {author} {\bibfnamefont {N.~K.}\ \bibnamefont
  {Madsen}}, \bibinfo {author} {\bibfnamefont {I.~H.}\ \bibnamefont
  {Godtliebsen}}, \bibinfo {author} {\bibfnamefont {S.~A.}\ \bibnamefont
  {Losilla}}, \ and\ \bibinfo {author} {\bibfnamefont {O.}~\bibnamefont
  {Christiansen}},\ }\href@noop {} {\bibfield  {journal} {\bibinfo  {journal}
  {J. Chem. Phys.}\ }\textbf {\bibinfo {volume} {148}},\ \bibinfo {pages}
  {024103} (\bibinfo {year} {2018})}\BibitemShut {NoStop}%
\bibitem [{\citenamefont {Baiardi}\ and\ \citenamefont
  {Reiher}(2019)}]{BaRe19}%
  \BibitemOpen
  \bibfield  {author} {\bibinfo {author} {\bibfnamefont {A.}~\bibnamefont
  {Baiardi}}\ and\ \bibinfo {author} {\bibfnamefont {M.}~\bibnamefont
  {Reiher}},\ }\href@noop {} {\bibfield  {journal} {\bibinfo  {journal} {J.
  Chem. Theory Comput.}\ }\textbf {\bibinfo {volume} {15}},\ \bibinfo {pages}
  {3481} (\bibinfo {year} {2019})}\BibitemShut {NoStop}%
\bibitem [{\citenamefont {Miller}\ \emph {et~al.}(1980)\citenamefont {Miller},
  \citenamefont {Handy},\ and\ \citenamefont {Adams}}]{MiHaAd80}%
  \BibitemOpen
  \bibfield  {author} {\bibinfo {author} {\bibfnamefont {W.~H.}\ \bibnamefont
  {Miller}}, \bibinfo {author} {\bibfnamefont {N.~C.}\ \bibnamefont {Handy}}, \
  and\ \bibinfo {author} {\bibfnamefont {J.~E.}\ \bibnamefont {Adams}},\
  }\href@noop {} {\bibfield  {journal} {\bibinfo  {journal} {J. Chem. Phys.}\
  }\textbf {\bibinfo {volume} {72}},\ \bibinfo {pages} {99} (\bibinfo {year}
  {1980})}\BibitemShut {NoStop}%
\bibitem [{\citenamefont {Bowman}\ \emph {et~al.}(2007)\citenamefont {Bowman},
  \citenamefont {Huang}, \citenamefont {Handy},\ and\ \citenamefont
  {Carter}}]{BoHuHaCa07}%
  \BibitemOpen
  \bibfield  {author} {\bibinfo {author} {\bibfnamefont {J.~M.}\ \bibnamefont
  {Bowman}}, \bibinfo {author} {\bibfnamefont {X.}~\bibnamefont {Huang}},
  \bibinfo {author} {\bibfnamefont {N.~C.}\ \bibnamefont {Handy}}, \ and\
  \bibinfo {author} {\bibfnamefont {S.}~\bibnamefont {Carter}},\ }\href@noop {}
  {\bibfield  {journal} {\bibinfo  {journal} {J. Phys. Chem. A}\ }\textbf
  {\bibinfo {volume} {111}},\ \bibinfo {pages} {7317} (\bibinfo {year}
  {2007})}\BibitemShut {NoStop}%
\bibitem [{\citenamefont {Lauvergnat}\ and\ \citenamefont
  {Nauts}(2014)}]{LAUVERGNAT201418}%
  \BibitemOpen
  \bibfield  {author} {\bibinfo {author} {\bibfnamefont {D.}~\bibnamefont
  {Lauvergnat}}\ and\ \bibinfo {author} {\bibfnamefont {A.}~\bibnamefont
  {Nauts}},\ }\href@noop {} {\ \textbf {\bibinfo {volume} {119}},\ \bibinfo
  {pages} {18 } (\bibinfo {year} {2014})}\BibitemShut {NoStop}%
\bibitem [{\citenamefont {Leforestier}(2012)}]{Leforestier:12a}%
  \BibitemOpen
  \bibfield  {author} {\bibinfo {author} {\bibfnamefont {C.}~\bibnamefont
  {Leforestier}},\ }\href@noop {} {\bibfield  {journal} {\bibinfo  {journal}
  {Phil. Trans. R. Soc. A}\ }\textbf {\bibinfo {volume} {370}},\ \bibinfo
  {pages} {2675} (\bibinfo {year} {2012})}\BibitemShut {NoStop}%
\bibitem [{\citenamefont {Heijmen}\ \emph {et~al.}(1997)\citenamefont
  {Heijmen}, \citenamefont {Korona}, \citenamefont {Moszynski}, \citenamefont
  {Wormer},\ and\ \citenamefont {van~der Avoird}}]{HeKoMoWoAv97}%
  \BibitemOpen
  \bibfield  {author} {\bibinfo {author} {\bibfnamefont {T.~G.~A.}\
  \bibnamefont {Heijmen}}, \bibinfo {author} {\bibfnamefont {T.}~\bibnamefont
  {Korona}}, \bibinfo {author} {\bibfnamefont {R.}~\bibnamefont {Moszynski}},
  \bibinfo {author} {\bibfnamefont {P.~E.~S.}\ \bibnamefont {Wormer}}, \ and\
  \bibinfo {author} {\bibfnamefont {A.}~\bibnamefont {van~der Avoird}},\
  }\href@noop {} {\bibfield  {journal} {\bibinfo  {journal} {J. Chem. Phys.}\
  }\textbf {\bibinfo {volume} {107}},\ \bibinfo {pages} {902} (\bibinfo {year}
  {1997})}\BibitemShut {NoStop}%
\bibitem [{\citenamefont {Miller}\ \emph {et~al.}(1999)\citenamefont {Miller},
  \citenamefont {Heijmen}, \citenamefont {Wormer}, \citenamefont {van~der
  Avoird},\ and\ \citenamefont {Moszynski}}]{MiHeWoAvMo99}%
  \BibitemOpen
  \bibfield  {author} {\bibinfo {author} {\bibfnamefont {R.~E.}\ \bibnamefont
  {Miller}}, \bibinfo {author} {\bibfnamefont {T.~G.~A.}\ \bibnamefont
  {Heijmen}}, \bibinfo {author} {\bibfnamefont {P.~E.~S.}\ \bibnamefont
  {Wormer}}, \bibinfo {author} {\bibfnamefont {A.}~\bibnamefont {van~der
  Avoird}}, \ and\ \bibinfo {author} {\bibfnamefont {R.}~\bibnamefont
  {Moszynski}},\ }\href@noop {} {\bibfield  {journal} {\bibinfo  {journal} {J.
  Chem. Phys.}\ }\textbf {\bibinfo {volume} {110}},\ \bibinfo {pages} {5651}
  (\bibinfo {year} {1999})}\BibitemShut {NoStop}%
\bibitem [{\citenamefont {Wangler}\ \emph {et~al.}(2003)\citenamefont
  {Wangler}, \citenamefont {Roth}, \citenamefont {Pak}, \citenamefont
  {Winnewisser}, \citenamefont {Wormer},\ and\ \citenamefont {van~der
  Avoird}}]{WaRoPaWiWoAv03}%
  \BibitemOpen
  \bibfield  {author} {\bibinfo {author} {\bibfnamefont {M.}~\bibnamefont
  {Wangler}}, \bibinfo {author} {\bibfnamefont {D.~A.}\ \bibnamefont {Roth}},
  \bibinfo {author} {\bibfnamefont {I.}~\bibnamefont {Pak}}, \bibinfo {author}
  {\bibfnamefont {G.}~\bibnamefont {Winnewisser}}, \bibinfo {author}
  {\bibfnamefont {P.~E.~S.}\ \bibnamefont {Wormer}}, \ and\ \bibinfo {author}
  {\bibfnamefont {A.}~\bibnamefont {van~der Avoird}},\ }\href {\doibase
  https://doi.org/10.1016/S0022-2852(03)00058-4} {\bibfield  {journal}
  {\bibinfo  {journal} {J. Mol. Spectrosc.}\ }\textbf {\bibinfo {volume}
  {222}},\ \bibinfo {pages} {109} (\bibinfo {year} {2003})}\BibitemShut
  {NoStop}%
\bibitem [{\citenamefont {Avila}\ and\ \citenamefont
  {M\'atyus}(2019)}]{AvMa19a}%
  \BibitemOpen
  \bibfield  {author} {\bibinfo {author} {\bibfnamefont {G.}~\bibnamefont
  {Avila}}\ and\ \bibinfo {author} {\bibfnamefont {E.}~\bibnamefont
  {M\'atyus}},\ }\href@noop {} {\bibfield  {journal} {\bibinfo  {journal} {J.
  Chem. Phys.}\ }\textbf {\bibinfo {volume} {150}},\ \bibinfo {pages} {174107}
  (\bibinfo {year} {2019})}\BibitemShut {NoStop}%
\bibitem [{\citenamefont {Adler}\ \emph {et~al.}(2007)\citenamefont {Adler},
  \citenamefont {Knizia},\ and\ \citenamefont {Werner}}]{AdKnWe07}%
  \BibitemOpen
  \bibfield  {author} {\bibinfo {author} {\bibfnamefont {T.~B.}\ \bibnamefont
  {Adler}}, \bibinfo {author} {\bibfnamefont {G.}~\bibnamefont {Knizia}}, \
  and\ \bibinfo {author} {\bibfnamefont {H.-J.}\ \bibnamefont {Werner}},\
  }\href@noop {} {\bibfield  {journal} {\bibinfo  {journal} {J. Chem. Phys.}\
  }\textbf {\bibinfo {volume} {127}},\ \bibinfo {pages} {221106} (\bibinfo
  {year} {2007})}\BibitemShut {NoStop}%
\bibitem [{\citenamefont {{Dunning, Jr.}}(1989)}]{Du89}%
  \BibitemOpen
  \bibfield  {author} {\bibinfo {author} {\bibfnamefont {T.~H.}\ \bibnamefont
  {{Dunning, Jr.}}},\ }\href@noop {} {\bibfield  {journal} {\bibinfo  {journal}
  {J. Chem. Phys.}\ }\textbf {\bibinfo {volume} {90}},\ \bibinfo {pages} {1007}
  (\bibinfo {year} {1989})}\BibitemShut {NoStop}%
\bibitem [{\citenamefont {Raghavachari}\ \emph {et~al.}(1989)\citenamefont
  {Raghavachari}, \citenamefont {Trucks}, \citenamefont {Pople},\ and\
  \citenamefont {Head-Gordon}}]{RaTrPoHG89}%
  \BibitemOpen
  \bibfield  {author} {\bibinfo {author} {\bibfnamefont {K.}~\bibnamefont
  {Raghavachari}}, \bibinfo {author} {\bibfnamefont {G.~W.}\ \bibnamefont
  {Trucks}}, \bibinfo {author} {\bibfnamefont {J.~A.}\ \bibnamefont {Pople}}, \
  and\ \bibinfo {author} {\bibfnamefont {M.}~\bibnamefont {Head-Gordon}},\
  }\href@noop {} {\bibfield  {journal} {\bibinfo  {journal} {Chem. Phys.
  Lett.}\ }\textbf {\bibinfo {volume} {157}},\ \bibinfo {pages} {479} (\bibinfo
  {year} {1989})}\BibitemShut {NoStop}%
\bibitem [{\citenamefont {K\'allay}\ and\ \citenamefont
  {Gauss}(2005)}]{KaGa05}%
  \BibitemOpen
  \bibfield  {author} {\bibinfo {author} {\bibfnamefont {M.}~\bibnamefont
  {K\'allay}}\ and\ \bibinfo {author} {\bibfnamefont {J.}~\bibnamefont
  {Gauss}},\ }\href@noop {} {\bibfield  {journal} {\bibinfo  {journal} {J.
  Chem. Phys.}\ }\textbf {\bibinfo {volume} {123}},\ \bibinfo {pages} {214105}
  (\bibinfo {year} {2005})}\BibitemShut {NoStop}%
\bibitem [{\citenamefont {Woon}\ and\ \citenamefont {{Dunning
  Jr.}}(1995)}]{WoDu95}%
  \BibitemOpen
  \bibfield  {author} {\bibinfo {author} {\bibfnamefont {D.~E.}\ \bibnamefont
  {Woon}}\ and\ \bibinfo {author} {\bibfnamefont {T.~H.}\ \bibnamefont
  {{Dunning Jr.}}},\ }\href@noop {} {\bibfield  {journal} {\bibinfo  {journal}
  {J. Chem. Phys.}\ }\textbf {\bibinfo {volume} {103}},\ \bibinfo {pages}
  {4572} (\bibinfo {year} {1995})}\BibitemShut {NoStop}%
\bibitem [{mol()}]{molpro}%
  \BibitemOpen
  \href@noop {} {}\bibinfo {note} {Molpro, version 2015.1, a package of
  \emph{ab initio} programs, H.-J. Werner, P. J. Knowles, G. Knizia, F. R.
  Manby, M. Sch\"utz, and others, see http://www.molpro.net.}\BibitemShut
  {Stop}%
\bibitem [{mrc()}]{mrcc}%
  \BibitemOpen
  \href@noop {} {}\bibinfo {note} {MRCC, a quantum chemical program suite
  written by M. K\'allay, Z. Rolik, I. Ladj\'anszki, L. Szegedy, B. Lad\'oczki,
  J. Csontos and B. Kornis, See also Z. Rolik, M. K\'allay, J. Chem. Phys.,
  2011, \textbf{135,} 104111, as well as: www.mrcc.hu.}\BibitemShut {Stop}%
\bibitem [{\citenamefont {Czak\'o}\ \emph {et~al.}(2014)\citenamefont
  {Czak\'o}, \citenamefont {Szab\'o},\ and\ \citenamefont
  {Telekes}}]{CzSzTe14}%
  \BibitemOpen
  \bibfield  {author} {\bibinfo {author} {\bibfnamefont {G.}~\bibnamefont
  {Czak\'o}}, \bibinfo {author} {\bibfnamefont {I.}~\bibnamefont {Szab\'o}}, \
  and\ \bibinfo {author} {\bibfnamefont {H.}~\bibnamefont {Telekes}},\
  }\href@noop {} {\bibfield  {journal} {\bibinfo  {journal} {J. Phys. Chem. A}\
  }\textbf {\bibinfo {volume} {118}},\ \bibinfo {pages} {646} (\bibinfo {year}
  {2014})}\BibitemShut {NoStop}%
\bibitem [{\citenamefont {Braams}\ and\ \citenamefont {Bowman}(2009)}]{BrBo09}%
  \BibitemOpen
  \bibfield  {author} {\bibinfo {author} {\bibfnamefont {B.~J.}\ \bibnamefont
  {Braams}}\ and\ \bibinfo {author} {\bibfnamefont {J.~M.}\ \bibnamefont
  {Bowman}},\ }\href@noop {} {\bibfield  {journal} {\bibinfo  {journal} {Int.
  Rev. Phys. Chem.}\ }\textbf {\bibinfo {volume} {28}},\ \bibinfo {pages} {577}
  (\bibinfo {year} {2009})}\BibitemShut {NoStop}%
\bibitem [{\citenamefont {Bowman}\ \emph {et~al.}(2011)\citenamefont {Bowman},
  \citenamefont {Czak\'o},\ and\ \citenamefont {Fu}}]{BoCzFu11}%
  \BibitemOpen
  \bibfield  {author} {\bibinfo {author} {\bibfnamefont {J.~M.}\ \bibnamefont
  {Bowman}}, \bibinfo {author} {\bibfnamefont {G.}~\bibnamefont {Czak\'o}}, \
  and\ \bibinfo {author} {\bibfnamefont {B.}~\bibnamefont {Fu}},\ }\href@noop
  {} {\bibfield  {journal} {\bibinfo  {journal} {Phys. Chem. Chem. Phys.}\
  }\textbf {\bibinfo {volume} {13}},\ \bibinfo {pages} {8094} (\bibinfo {year}
  {2011})}\BibitemShut {NoStop}%
\bibitem [{\citenamefont {M\'atyus}\ \emph {et~al.}(2009)\citenamefont
  {M\'atyus}, \citenamefont {Czak\'o},\ and\ \citenamefont
  {Cs\'asz\'ar}}]{MaCzCs09}%
  \BibitemOpen
  \bibfield  {author} {\bibinfo {author} {\bibfnamefont {E.}~\bibnamefont
  {M\'atyus}}, \bibinfo {author} {\bibfnamefont {G.}~\bibnamefont {Czak\'o}}, \
  and\ \bibinfo {author} {\bibfnamefont {A.~G.}\ \bibnamefont {Cs\'asz\'ar}},\
  }\href@noop {} {\bibfield  {journal} {\bibinfo  {journal} {J. Chem. Phys.}\
  }\textbf {\bibinfo {volume} {130}},\ \bibinfo {pages} {134112} (\bibinfo
  {year} {2009})}\BibitemShut {NoStop}%
\bibitem [{\citenamefont {F\'abri}\ \emph {et~al.}(2011)\citenamefont
  {F\'abri}, \citenamefont {M\'atyus},\ and\ \citenamefont
  {Cs\'asz\'ar}}]{FaMaCs11}%
  \BibitemOpen
  \bibfield  {author} {\bibinfo {author} {\bibfnamefont {C.}~\bibnamefont
  {F\'abri}}, \bibinfo {author} {\bibfnamefont {E.}~\bibnamefont {M\'atyus}}, \
  and\ \bibinfo {author} {\bibfnamefont {A.~G.}\ \bibnamefont {Cs\'asz\'ar}},\
  }\href@noop {} {\bibfield  {journal} {\bibinfo  {journal} {J. Chem. Phys.}\
  }\textbf {\bibinfo {volume} {134}},\ \bibinfo {pages} {074105} (\bibinfo
  {year} {2011})}\BibitemShut {NoStop}%
\bibitem [{\citenamefont {Wang}\ and\ \citenamefont
  {Carrington~Jr}(2014)}]{WaCa14}%
  \BibitemOpen
  \bibfield  {author} {\bibinfo {author} {\bibfnamefont {X.-G.}\ \bibnamefont
  {Wang}}\ and\ \bibinfo {author} {\bibfnamefont {T.}~\bibnamefont
  {Carrington~Jr}},\ }\href@noop {} {\bibfield  {journal} {\bibinfo  {journal}
  {J. Chem. Phys.}\ }\textbf {\bibinfo {volume} {141}},\ \bibinfo {pages}
  {154106} (\bibinfo {year} {2014})}\BibitemShut {NoStop}%
\bibitem [{\citenamefont {Coursey}\ \emph {et~al.}(2015)\citenamefont
  {Coursey}, \citenamefont {Schwab}, \citenamefont {Tsai},\ and\ \citenamefont
  {Dragoset}}]{NIST}%
  \BibitemOpen
  \bibfield  {author} {\bibinfo {author} {\bibfnamefont {J.~S.}\ \bibnamefont
  {Coursey}}, \bibinfo {author} {\bibfnamefont {D.~J.}\ \bibnamefont {Schwab}},
  \bibinfo {author} {\bibfnamefont {J.~J.}\ \bibnamefont {Tsai}}, \ and\
  \bibinfo {author} {\bibfnamefont {R.~A.}\ \bibnamefont {Dragoset}},\
  }\href@noop {} {}\bibinfo {howpublished} {{Atomic Weights and Isotopic
  Compositions (version 4.1): {\tt{http://physics.nist.gov/Comp}} [last
  accessed on 12 May 2018]. National Institute of Standards and Technology,
  Gaithersburg, MD.}} (\bibinfo {year} {2015})\BibitemShut {NoStop}%
\bibitem [{\citenamefont {Schwenke}\ and\ \citenamefont
  {Partridge}(2001)}]{ScPa01}%
  \BibitemOpen
  \bibfield  {author} {\bibinfo {author} {\bibfnamefont {D.~W.}\ \bibnamefont
  {Schwenke}}\ and\ \bibinfo {author} {\bibfnamefont {H.}~\bibnamefont
  {Partridge}},\ }\href@noop {} {\bibfield  {journal} {\bibinfo  {journal}
  {Spectrochim. Acta}\ }\textbf {\bibinfo {volume} {57}},\ \bibinfo {pages}
  {887} (\bibinfo {year} {2001})}\BibitemShut {NoStop}%
\bibitem [{\citenamefont {Nikitin}\ \emph {et~al.}(2011)\citenamefont
  {Nikitin}, \citenamefont {Rey},\ and\ \citenamefont {Tyuterev}}]{NiReTy11}%
  \BibitemOpen
  \bibfield  {author} {\bibinfo {author} {\bibfnamefont {A.~V.}\ \bibnamefont
  {Nikitin}}, \bibinfo {author} {\bibfnamefont {M.}~\bibnamefont {Rey}}, \ and\
  \bibinfo {author} {\bibfnamefont {V.~G.}\ \bibnamefont {Tyuterev}},\
  }\href@noop {} {\bibfield  {journal} {\bibinfo  {journal} {Chem. Phys.
  Lett.}\ }\textbf {\bibinfo {volume} {501}},\ \bibinfo {pages} {179} (\bibinfo
  {year} {2011})}\BibitemShut {NoStop}%
\bibitem [{\citenamefont {Ferenc}\ and\ \citenamefont
  {M\'atyus}(2019)}]{FeMa19}%
  \BibitemOpen
  \bibfield  {author} {\bibinfo {author} {\bibfnamefont {D.}~\bibnamefont
  {Ferenc}}\ and\ \bibinfo {author} {\bibfnamefont {E.}~\bibnamefont
  {M\'atyus}},\ }\href@noop {} {\bibfield  {journal} {\bibinfo  {journal} {Mol.
  Phys.}\ }\textbf {\bibinfo {volume} {117}},\ \bibinfo {pages} {1694}
  (\bibinfo {year} {2019})}\BibitemShut {NoStop}%
\bibitem [{\citenamefont {Wei}\ and\ \citenamefont {{Carrington,
  Jr.}}(1992)}]{WeCa92}%
  \BibitemOpen
  \bibfield  {author} {\bibinfo {author} {\bibfnamefont {H.}~\bibnamefont
  {Wei}}\ and\ \bibinfo {author} {\bibfnamefont {T.}~\bibnamefont {{Carrington,
  Jr.}}},\ }\href@noop {} {\bibfield  {journal} {\bibinfo  {journal} {J. Chem.
  Phys.}\ }\textbf {\bibinfo {volume} {97}},\ \bibinfo {pages} {3029} (\bibinfo
  {year} {1992})}\BibitemShut {NoStop}%
\bibitem [{\citenamefont {Johnson}\ and\ \citenamefont
  {Reinhardt}(1986)}]{doi:10.1063/1.451775}%
  \BibitemOpen
  \bibfield  {author} {\bibinfo {author} {\bibfnamefont {B.~R.}\ \bibnamefont
  {Johnson}}\ and\ \bibinfo {author} {\bibfnamefont {W.~P.}\ \bibnamefont
  {Reinhardt}},\ }\href@noop {} {\bibfield  {journal} {\bibinfo  {journal} {J.
  Chem. Phys.}\ }\textbf {\bibinfo {volume} {85}},\ \bibinfo {pages} {4538}
  (\bibinfo {year} {1986})}\BibitemShut {NoStop}%
\bibitem [{\citenamefont {Tennyson}\ and\ \citenamefont
  {Sutcliffe}(1982)}]{doi:10.1063/1.444316}%
  \BibitemOpen
  \bibfield  {author} {\bibinfo {author} {\bibfnamefont {J.}~\bibnamefont
  {Tennyson}}\ and\ \bibinfo {author} {\bibfnamefont {B.~T.}\ \bibnamefont
  {Sutcliffe}},\ }\href@noop {} {\bibfield  {journal} {\bibinfo  {journal} {J.
  Chem. Phys.}\ }\textbf {\bibinfo {volume} {77}},\ \bibinfo {pages} {4061}
  (\bibinfo {year} {1982})}\BibitemShut {NoStop}%
\bibitem [{\citenamefont {Schiffel}\ and\ \citenamefont
  {Manthe}(2010)}]{SCHIFFEL2010118}%
  \BibitemOpen
  \bibfield  {author} {\bibinfo {author} {\bibfnamefont {G.}~\bibnamefont
  {Schiffel}}\ and\ \bibinfo {author} {\bibfnamefont {U.}~\bibnamefont
  {Manthe}},\ }\href@noop {} {\bibfield  {journal} {\bibinfo  {journal}
  {Chemical Physics}\ }\textbf {\bibinfo {volume} {374}},\ \bibinfo {pages}
  {118 } (\bibinfo {year} {2010})}\BibitemShut {NoStop}%
\end{thebibliography}

\end{document}